\documentclass[aps,preprint,floatfix]{revtex4}%
\usepackage{amsfonts}
\usepackage{amsmath}
\usepackage{amssymb}
\usepackage{graphicx}%
\providecommand{\U}[1]{\protect\rule{.1in}{.1in}}
\begin{document}
\title{Generalized Diffusion}
\author{James F. Lutsko}
\affiliation{Physics Department CP 231, Universit\'e Libre de Bruxelles, 1050 - Bruxelles, Belgium}
\email{jlutsko@ulb.ac.be}
\homepage{http://www.lutsko.com}
\author{Jean Pierre Boon}
\affiliation{Physics Department CP 231, Universit\'e Libre de Bruxelles, 1050 - Bruxelles, Belgium}
\email{jpboon@ulb.ac.be}
\homepage{http://poseidon.ulb.ac.be}


\begin{abstract}
The Fokker-Planck equation for the probability $f(r,t)$ to find a random
walker at position $r$ at time $t$ is derived for the case that the the
probability to make jumps depends nonlinearly on $f(r,t)$. The result is a
generalized form of the classical Fokker-Planck equation where the effects of
drift, due to a violation of detailed balance, and of external fields are also
considered. It is shown that in the absence of drift and external fields a
scaling solution, describing anomalous diffusion, is only possible if the
nonlinearity in the jump probability is of the power law type ($\sim f^{\eta
}(r,t)$), in which case the generalized Fokker-Planck equation reduces to the
well-known Porous Media equation. Monte-Carlo simulations are shown to confirm
the theoretical results.

\bigskip


\end{abstract}
\date{\today}
\maketitle

\section{Introduction}

Random walks are typically characterized by the probability to find a walker
at some position $r$ at some time $t$, $f\left(  r,t\right)  $ . This could
equally well be the concentration of walkers at a given space-time location in
the event that there is an ensemble of independent walkers. In the classical
case of an unbiased, discrete-time random walk on a lattice, it was first
shown by Einstein that at length and time scales large compared to the lattice
spacing and the time step, respectively, the distribution satisfies the
Fokker-Planck equation%
\begin{equation}
\frac{\partial}{\partial t}f\left(  r,t\right)  =D\frac{\partial^{2}}{\partial
r^{2}}f\left(  r,t\right)  \,, \label{1}%
\end{equation}
which is the classical diffusion equation with diffusion constant $D$
\cite{Einstein}$.$ The diffusion constant can be expressed in terms of the
microscopic dynamics of the problem, namely the probability for the walker to
take a step, the lattice spacing and the time step. It is obvious by
inspection that the diffusion equation admits of normalized scaling solutions
of the form $f\left(  r,t\right)  =t^{-1/2}\phi\left(  r^{2}/t\right)  $ which
immediately implies typical diffusive scaling of the second moment,
$\left\langle r^{2}\right\rangle _{t}=2Dt$, which is the Einstein relation.
However, there are many systems observed in nature where it seems natural to
use the language of diffusion, but for which the mean-squared displacement
scales as something other than linearly with time. In order to describe such
systems, the functional form of eq.(\ref{1}) is often generalized so as to
allow for other types of scalings. Two popular methods are (i) the replacement
of the second derivative on the right by a fractional derivative giving the
Fractional Fokker Planck equation \cite{Metzler,Metzler_Review} (FFPE) and (ii) replacing
$f\left(  r,t\right)  $ on the right by $f^{\alpha}\left(  r,t\right)  $
giving the Porous Media equation \cite{muskat, plastino,tsallis} (PME). The
FFPE can be understood as being the continuum limit of a continuous time
random walk in which the waiting times obey the Mittag-Leffler distribution
\cite{Goychuk} (a generalization of the Poisson distribution in which the
probability for a jump decays algebraically with time for long times). It is
therefore possible in this case to relate the mathematical formalism (the
FFPE) to a microscopic description (power-law-distributed waiting times). The
purpose of this paper is to describe a similar class of microscopic dynamics
for which the PME arises naturally as the corresponding Fokker-Planck equation.

There have been several attempts to provide some dynamical context for the
PME. Abe and Thurner \cite{AbeThurner} attempted to generalize the classical
derivation of Einstein by introducing the concept of escort probabilities into
the master equation for a random walk. Aside from the ad hoc nature of the
generalization, the result is the PME plus an additional term which is not
well-behaved in the long-time limit. Several authors have described the
relation of the PME to a continuous time random walk. In particular, Curado
and Nobre \cite{CuradoNobre} show that the PME arises from a continuous time
random walk in which the transition rates, which are constants in the
classical random walk, depend on some power of the distribution. Borland
\cite{Borland} and Anteneodo and Tsallis \cite{AnteneodoTsallis} discuss the
fact that the PME corresponds to a Langevin equation with multiplicative noise
but, given the equivalence of the Fokker-Planck and Langevin descriptions,
this is just another way of writing the same result. Lutsko and Boon
\cite{LutskoBoon} show that an assumption of nonlinear response in an ordinary
fluid leads to the PME, but with no indication of the origin of the nonlinear
response. Another approach based on generalizing the cumulant expansion of the
intermediate scattering function leads to a somewhat different generalization
of classical diffusion \cite{BoonLutsko}.

Consider a discrete time random walk on a one dimensional lattice under the
condition that the probability that the walker makes a jump from one lattice
site to another depends on the concentration of walkers everywhere on the
lattice. In this way, we generalize and extend previous models in several
ways. First, we allow for jumps of arbitrary length and with asymmetric
probabilities so that detailed balance is violated and an intrinsic drift is
generated. Second, we start with a discrete-time model rather than the
continuous time random walk which, combined with the drift, leads to
additional terms in the Fokker-Planck equation. The continuous time random
walk is a special limit of our formulation. Third, we do not assume a priori
power-law nonlinearities as has commonly been the case. We allow for a quite
general form of nonlinear dependence of the jump-probabilities on the local
distributions and we show that the resulting Fokker-Planck equation only
admits self-similar, i.e. scaling, solutions if the nonlinearities take the form
of power laws. Thus we derive the existence of power-law nonlinearities rather
than impose them. Finally, we also consider the effect of an external
field. A preliminary discussion of these results has recently appeared\cite{BoonLutsko_EPL_2007}.

In the next section, we start from the master equation and use a multiscale
expansion to derive the Fokker-Planck equation. The modifications necessary to
take into account the action of an external field are also discussed. In
section III, we explore the properties of the generalized diffusion equation.
In particular, we show that self-similar solutions are only possible under
conditions that reduce our equation to the PME. We also present numerical
results which demonstrate the importance of the non-standard terms occuring in
the generalized diffusion equation and showing, in particular, the effect of
breaking detailed balance. The last section gives our conclusions.

\section{Derivation of the generalized Fokker-Planck equation}

\subsection{The master equation}

Consider a walker on a lattice whose sites are labeled by a discrete index,
$l$. A classical random walk is characterized by a set of probabilities
$\left\{  p_{j}\right\}  $ which give the likelihood for a jump of $j$ lattice
sites ( $j>0$ corresponds to jumps to the right, $j<0$ to jumps to the left).
An individual walker is characterized by the probability to be at site $l$ at
time step $i$, $\widehat{f_{l}}\left(  i\right)  $. Equivalently, one could
imagine a population of independent walkers which all start from the same
site, in which case $\widehat{f}_{l}(i)$ would be the concentration of walkers
at site $l$ at time step $i$. If the walk is symmetric, $p_{-j}=p_{j}$, the
walker exhibits diffusive behavior whereas asymmetric probabilities give rise
to diffusion superposed on a systematic drift.

In the present case, we generalize this picture by considering that the jump
probability is a function of the occupation probability (or the concentration
of particles) on the lattice. Consequently the probability to make a jump of
length $j$ from site $l$ will depend on the concentration at site $l$ at time
step $i$, $\widehat{f}_{l}\left(  i\right)$, and on the concentration at the end point
of the jump, $\widehat{f}_{l+j}\left(  i\right)  $. It is convenient to
introduce the more general notation whereby the probability to jump from site
$l$ to site $k$ at time $t$ is $P\left(  l\rightarrow k;t\right)  $ so that
the distribution obeys the master equation%
\begin{equation}
\widehat{f}_{l}\left(  i+1\right)  =\widehat{f}_{l}\left(  i\right)
+\sum_{k=-\infty}^{\infty}\left[  \widehat{f}_{k}\left(  i\right)  P\left(
k\rightarrow l;i\right)  -\widehat{f}_{l}\left(  i\right)  P\left(
l\rightarrow k;i\right)  \right]  \label{discrete}%
\end{equation}
where the first term on the right is the increase in population due to walkers
jumping to site $l$ from all other sites, $k$, whereas the second term is the
loss due to walkers leaving site $l$ to go to site $k$. In the classical case,
the jump probabilities take values drawn from a prescribed distribution, i.e.
$P\left(  k\rightarrow l;t\right)  =p_{l-k}$ \footnote{The simplest case being
$p_{(k+1)-k}=p_{(k-1)-k}=\frac{1}{2}$.}. Here, we make the specific
generalization that the probabilities have the form%
\begin{equation}
P\left(  k\rightarrow l;i\right)  =p_{l-k}F\left(  \widehat{f}_{k}\left(
i\right)  ,\widehat{f}_{l}\left(  i\right)  \right)  \label{jumps}%
\end{equation}
for some, as yet unspecified, function $F(x,y)$. Note that the probabilities
must satisfy the obvious normalization%
\begin{equation}
1=\sum_{k}P\left(  l\rightarrow k;t\right)  .
\end{equation}
This, together with the requirement that the probabilities be bounded, $0\leq
P\left(  k\rightarrow l;t\right)  \leq1$, places restrictions on the form of
$F\left(  x,y\right)  $.

\subsection{Smoothing}

The goal is to examine the distribution on length and time scales that are
large compared to the lattice spacing $\delta r$ and time step $\delta t$. On
these scales, it is expected that the distribution can be approximated by a
continuous function.To formalize this, a smoothed version of the distribution
is defined by%
\begin{equation}
f\left(  r,t\right)  =\sum_{l,i}G\left(  r-l\delta r,t-i\delta t\right)
\widehat{f}_{l}\left(  i\right)
\end{equation}
where the sum extends over all values of the indices and the function $G(r,t)$
is assumed to be localized near the point $r=0,t=0$. For example, the
smoothing function could be a product of gaussians,%
\begin{equation}
G\left(  r,t\right)  =\frac{1}{2\pi\sqrt{\sigma_{r}\sigma_{t}}}\exp\left(
-\frac{r^{2}}{2\sigma_{r}}\right)  \exp\left(  -\frac{t^{2}}{2\sigma_{t}%
}\right)  .
\end{equation}
In general, the length and time scales associated with the smoothing can be as
small as those of the random walk model. In the following, it makes no
difference as long as both are small compared to the scale of typical
variations in the distribution function. In general, we assume that, as in
this example, there are scales such as $\sigma_{r}$ and $\sigma_{t}$ that
characterize the range of the smoothing and henceforth that these scales are
similar to the lattice spacing and time step,
\begin{equation}
1\lesssim\delta r/\sqrt{\sigma_{r}},\delta t/\sqrt{\sigma_{t}}.
\end{equation}

It will be necessary to also define the inverse transformation. To that end,
notice that%
\begin{equation}
f\left(  k\delta r,j\delta t\right)  =\sum_{l,i}G\left(  k\delta r-l\delta
r,j\delta t-i\delta t\right)  \widehat{f}_{l}\left(  i\right)
\end{equation}
and assume that this relation is invertible so that
\begin{equation}
\widehat{f}_{l}\left(  i\right)  =\sum_{k,j}G^{-1}\left(  k\delta r-l\delta
r,j\delta t-i\delta t\right)  f\left(  k\delta r,j\delta t\right)  .
\end{equation}
This expression can be developed in a Taylor expansion. Assuming that the
smoothing function and its inverse are even functions of their arguments, as
is natural, then%
\begin{align}
\widehat{f}_{l}\left(  i\right)   &  =f\left(  l\delta r,i\delta t\right)
\sum_{k,j}G^{-1}\left(  k\delta r-l\delta r,j\delta t-i\delta t\right)
\nonumber\\
&  +\frac{1}{2}\left(  \delta r\right)  ^{2}\frac{\partial^{2}f\left(  l\delta
r,i\delta t\right)  }{\partial r^{2}}\sum_{k,j}\left(  k-l\right)  ^{2}%
G^{-1}\left(  k\delta r-l\delta r,j\delta t-i\delta t\right) \nonumber\\
&  +\frac{1}{2}\left(  \delta t\right)  ^{2}\frac{\partial^{2}f\left(  l\delta
r,i\delta t\right)  }{\partial t^{2}}\sum_{k,j}\left(  j-i\right)  ^{2}%
G^{-1}\left(  k\delta r-l\delta r,j\delta t-i\delta t\right)  \,+\, ...
\end{align}
It is easy to see that if $G\left(  k\delta r-l\delta r,j\delta t-i\delta
t\right)  $ is normalized, then so is the inverse function. The sums then
characterize the width in space and time, respectively, of the inverse
smoothing functions which will be of the same order of magnitude as that of
the actual smoothing functions. So, we have that%
\begin{equation}
\widehat{f}_{l}\left(  i\right)  =f\left(  l\delta r,i\delta t\right)
+\gamma_{r}\sigma_{r}\frac{\partial^{2}f\left(  l\delta r,i\delta t\right)
}{\partial r^{2}}+\gamma_{t}\sigma_{t}\frac{\partial^{2}f\left(  l\delta
r,i\delta t\right)  }{\partial t^{2}}+... \label{expand}%
\end{equation}
for some dimensionless constants $\gamma_{r},\gamma_{t}$ which are of order
unity. Note that this expansion makes sense, as does the whole smoothing
procedure, provided the gradients of the distribution are small over the
scales $\sqrt{\sigma_{r}},\sqrt{\sigma_{t}}$.

\subsection{Expansion of the master equation}

In the limit of classical diffusion, when the transition probabilities take
values from a given distribution, one could simply multiply the master
equation by $G\left(  r-l\delta r,t-i\delta t\right)  $ and sum to get the
exact master equation for the smoothed distribution,%
\begin{equation}
f\left(  r,t+\delta t\right)  =f\left(  r,t\right)  +\sum_{m=-\infty}^{\infty
}\left[  f\left(  r-k\delta r,t\right)  p_{m}-f\left(  r,t\right)
p_{-m}\right]
\end{equation}
However, the nonlinearities of the generalized model do not permit this.
Instead, eq.(\ref{expand}) is used to get
\begin{align}
&  f\left(  r,t+\delta t\right)  +\gamma_{r}\sigma_{r}\frac{\partial
^{2}f\left(  r,t+\delta t\right)  }{\partial r^{2}}+\gamma_{t}\sigma_{t}%
\frac{\partial^{2}f\left(  r,t+\delta t\right)  }{\partial t^{2}%
}+...\nonumber\label{expand1}\\
&  =f\left(  r,t\right)  +\gamma_{r}\sigma_{r}\frac{\partial^{2}f\left(
r,t\right)  }{\partial r^{2}}+\gamma_{t}\sigma_{t}\frac{\partial^{2}f\left(
r,t\right)  }{\partial t^{2}}+...\nonumber\\
&  +\sum_{m=-\infty}^{\infty}\left[  f\left(  r-m\delta r,t\right)  F\left(
f\left(  r-m\delta r,t\right)  ,f\left(  r,t\right)  \right)  p_{m}-f\left(
r,t\right)  F\left(  f\left(  r,t\right)  ,f\left(  r-m\delta r,t\right)
\right)  p_{-m}\right] \nonumber\\
&  +\gamma_{r}\sigma_{r}\sum_{m=-\infty}^{\infty} [ \frac{\partial^{2}f\left(
r-m\delta r,t\right)  }{\partial r^{2}}F\left(  f\left(  r-m\delta r,t\right)
,f\left(  r,t\right)  \right)  p_{m}\nonumber\\
&  \;\;\;\;\;\;\;\;\;\;\;\;\;\;\;\;\;\; - \frac{\partial^{2}f\left(
r,t\right)  }{\partial r^{2}}F\left(  f\left(  r,t\right)  ,f\left(  r-m\delta
r,t\right)  \right)  p_{-m} ] + ...
\end{align}
where we have only explicitly written one of several terms in the sum
proportional to $\sigma_{r}$ (and none of the terms proportional to
$\sigma_{t}$). The reason is that we will now further expand the distribution
so as to give a superficially local expression. Then, it is found that the
terms proportional to $\sigma_{r}$ and $\sigma_{t}$ only contribute to third
order in the gradients, so that we have%
\begin{align}
&  \delta t\frac{\partial f\left(  r,t\right)  }{\partial t}+\frac{1}%
{2}\left(  \delta t\right)  ^{2}\frac{\partial^{2}f\left(  r,t\right)
}{\partial t^{2}}\nonumber\\
&  =-\delta r\frac{\partial f\left(  r,t\right)  }{\partial r}\sum_{m=-\infty
}^{\infty}mp_{m}\left[  \frac{\partial xF\left(  x,y\right)  }{\partial
x}+\frac{\partial xF\left(  x,y\right)  }{\partial y}\right]  _{f}\nonumber\\
&  +\frac{1}{2}\left(  \delta r\right)  ^{2}\frac{\partial^{2}f\left(
r,t\right)  }{\partial r^{2}}\sum_{m=-\infty}^{\infty}m^{2}p_{m}\left[
\frac{\partial xF\left(  x,y\right)  }{\partial x}-\frac{\partial xF\left(
x,y\right)  }{\partial y}\right]  _{f}\nonumber\\
&  +\frac{1}{2}\left(  \delta r\right)  ^{2}\left(  \frac{\partial f\left(
r,t\right)  }{\partial r}\right)  ^{2}\sum_{m=-\infty}^{\infty}m^{2}%
p_{m}\left[  \frac{\partial^{2}xF\left(  x,y\right)  }{\partial x^{2}}%
-\frac{\partial^{2}xF\left(  x,y\right)  }{\partial y^{2}}\right]
_{f}\nonumber\\
&  + \mathcal{O} \left(  \sigma_{r}^{3/2}\frac{\partial^{3}f}{\partial r^{3}%
},...\right)
\end{align}
where we have used the assumption that $\delta r<\sqrt{\sigma_{r}}$ to replace
$\delta r$ by $\sigma_{r}$ in the error estimate. A compact notation has also
been introduced whereby%
\begin{equation}
\left(  \frac{\partial xF\left(  x,y\right)  }{\partial x}\right)
_{f}=\left(  \frac{\partial xF\left(  x,y\right)  }{\partial x}\right)
_{x=f\left(  r,t\right)  ,y=f\left(  r,t\right)  }.
\end{equation}

\subsection{Multiple time scales}

We could simply truncate the expansion obtained so far on the grounds that the
gradients are small over the scale of the smoothing (i.e. small over the scale
of a few lattice spacings) but this is unsatisfactory on both physical and
mathematical grounds. Physically, the resulting equation does not reduce to
the diffusion equation in the appropriate limit of $F\left(  x,y\right)  =1$.
Mathematically, this results in a second order equation in time whereas the
exact master equation is clearly first order in time:\ knowledge of the
distribution at time step $i$ is sufficient to calculate it at all future time
steps. These problems are not unrelated:\ both are due to the fact that
changes in the distribution in time are driven by spatial gradients so that in
some sense derivatives in time and in space are interchangeable. Ideally, we
would like to say that the first order spatial gradients drive the first order
time derivative, the second order gradients second order time derivatives,
etc. However, we cannot simply equate these different terms separately as
there is only one distribution and it can only satisfy one equation. The
solution is to generalize the distribution to have many different, but
related, time dependencies that can be satisfied at different length scales.
This leads to the method of multiple time scales.

To separate the different length and time scales in the problem, first define
a length scale $\ell$ over which the relative variation of the distribution is
of order one,
\begin{equation}
\frac{1}{f}\,\frac{\partial f}{\partial r/\ell}\sim1\,,\;\;\;\mbox{or}\;\;\;\ell
\,\frac{\partial\ln f}{\partial r}\sim1.
\end{equation}
Then, a small parameter $\epsilon=\delta r/\ell$ is defined which quantifies
the notion that the derivative of the distribution is small over the
length-scale of the smoothing function (which we assume is a few lattice
spacings so that $\delta r\sim\sqrt{\sigma_{r}}$). A parameter $\tau$ is
introduced by defining $\delta t=\epsilon\tau$ and dimensionless variables
$z=r/\ell$ and $s=t/\tau$ are used to write the master equation as%
\begin{align}
&  \epsilon\frac{\partial f\left(  z,s\right)  }{\partial s}+\frac{1}%
{2}\epsilon^{2}\frac{\partial^{2}f\left(  z,s\right)  }{\partial s^{2}%
}\nonumber\\
&  =-\epsilon\frac{\partial f\left(  z,s\right)  }{\partial z}J_{1}\left[
\frac{\partial xF\left(  x,y\right)  }{\partial x}+\frac{\partial xF\left(
x,y\right)  }{\partial y}\right]  _{f}\nonumber\\
&  +\frac{1}{2}\epsilon^{2}\frac{\partial^{2}f\left(  z,s\right)  }{\partial
z^{2}}J_{2}\left[  \frac{\partial xF\left(  x,y\right)  }{\partial x}%
-\frac{\partial xF\left(  x,y\right)  }{\partial y}\right]  _{f}\nonumber\\
&  +\frac{1}{2}\epsilon^{2}\left(  \frac{\partial f\left(  z,s\right)
}{\partial z}\right)  ^{2}J_{2}\left[  \frac{\partial^{2}xF\left(  x,y\right)
}{\partial x^{2}}-\frac{\partial^{2}xF\left(  x,y\right)  }{\partial y^{2}%
}\right]  _{f}  +\cal{O}\left(  \epsilon^{3}\right)
\end{align}
where $J_{n}=\sum_{m}m^{n}p_{m}$. Additional time scales are now introduced by
generalizing the distribution to a function of many time variables $f\left(
z,s\right)  \rightarrow f\left(  z,s_{0},s_{1},...\right)  $ where the
connection between this generalized function and the actual distribution is
that $f\left(  z,s\right)  =f\left(  z,s,\epsilon s,\epsilon^{2}s,...\right)
$. Thus, the time derivatives must be replaced by%
\begin{equation}
\frac{\partial}{\partial s}=\frac{\partial s_{0}}{\partial s}\frac{\partial
}{\partial s_{0}}+\frac{\partial s_{1}}{\partial s}\frac{\partial}{\partial
s_{1}}+...=\frac{\partial}{\partial s_{0}}+\epsilon\frac{\partial}{\partial
s_{1}}+...
\end{equation}
We can now demand that the terms cancel at each order in $\epsilon$ since this
just defines the dependence of the distribution on the various time-scales.
The first two orders in $\epsilon$ gives%
\begin{align}
\frac{\partial f}{\partial s_{0}}  &  =-\frac{\partial f}{\partial z}%
J_{1}\left[  \frac{\partial xF\left(  x,y\right)  }{\partial x}+\frac{\partial
xF\left(  x,y\right)  }{\partial y}\right]  _{f}\nonumber\\
\frac{\partial f}{\partial s_{1}}+\frac{1}{2}\frac{\partial^{2}f}{\partial
s_{0}^{2}}  &  =\frac{1}{2}\frac{\partial^{2}f}{\partial z^{2}}J_{2}\left[
\frac{\partial xF\left(  x,y\right)  }{\partial x}-\frac{\partial xF\left(
x,y\right)  }{\partial y}\right]  _{f}\nonumber\\
&  +\frac{1}{2}\left(  \frac{\partial f}{\partial z}\right)  ^{2}J_{2}\left[
\frac{\partial^{2}xF\left(  x,y\right)  }{\partial x^{2}}-\frac{\partial
^{2}xF\left(  x,y\right)  }{\partial y^{2}}\right]  _{f}%
\end{align}
Now it is clear that the first equation can be used to rewrite the second
derivative with respect to $s_{0}$ in terms of spatial gradients,%
\begin{align}
\frac{\partial^{2}f}{\partial s_{0}^{2}}  &  =\frac{\partial}{\partial s_{0}%
}\left(  -\frac{\partial f}{\partial z}J_{1}\left[  \frac{\partial xF\left(
x,y\right)  }{\partial x}+\frac{\partial xF\left(  x,y\right)  }{\partial
y}\right]  _{f}\right) \nonumber\\
&  =J_{1}^{2}\frac{\partial}{\partial z}\left(  \frac{\partial xF\left(
x,x\right)  }{\partial x}\right)  _{f}^{2}\frac{\partial f}{\partial z}%
\end{align}
giving%
\begin{align}
\frac{\partial f}{\partial s_{0}}  &  =-\frac{\partial f}{\partial z}%
J_{1}\left[  \frac{\partial xF\left(  x,y\right)  }{\partial x}+\frac{\partial
xF\left(  x,y\right)  }{\partial y}\right]  _{f}\nonumber\\
\frac{\partial f}{\partial s_{1}}  &  =\frac{1}{2}\frac{\partial^{2}%
f}{\partial z^{2}}J_{2}\left[  \frac{\partial xF\left(  x,y\right)  }{\partial
x}-\frac{\partial xF\left(  x,y\right)  }{\partial y}\right]  _{f}\nonumber\\
&  +\frac{1}{2}\left(  \frac{\partial f}{\partial z}\right)  ^{2}J_{2}\left[
\frac{\partial^{2}xF\left(  x,y\right)  }{\partial x^{2}}-\frac{\partial
^{2}xF\left(  x,y\right)  }{\partial y^{2}}\right]  _{f}\nonumber\\
&  -\frac{1}{2}J_{1}^{2}\frac{\partial}{\partial z}\left(  \frac{\partial
xF\left(  x,x\right)  }{\partial x}\right)  _{f}^{2}\frac{\partial f}{\partial
z}%
\end{align}
Summing gives the desired result,%
\begin{align}
\frac{\partial f}{\partial s}  &  =-\frac{\partial f}{\partial z}J_{1}\left[
\frac{\partial xF\left(  x,y\right)  }{\partial x}+\frac{\partial xF\left(
x,y\right)  }{\partial y}\right]  _{f}\nonumber\\
&  +\frac{1}{2}\frac{\partial^{2}f}{\partial z^{2}}J_{2}\left[  \frac{\partial
xF\left(  x,y\right)  }{\partial x}-\frac{\partial xF\left(  x,y\right)
}{\partial y}\right]  _{f}\nonumber\\
&  +\frac{1}{2}\left(  \frac{\partial f}{\partial z}\right)  ^{2}J_{2}\left[
\frac{\partial^{2}xF\left(  x,y\right)  }{\partial x^{2}}-\frac{\partial
^{2}xF\left(  x,y\right)  }{\partial y^{2}}\right]  _{f}\nonumber\\
&  -\frac{1}{2}J_{1}^{2}\frac{\partial}{\partial z}\left(  \frac{\partial
xF\left(  x,x\right)  }{\partial x}\right)  _{f}^{2}\frac{\partial f}{\partial
z}  +\mathcal{O}\left(  \epsilon^{2}\right)
\end{align}
In terms of the original variables, this reads%
\begin{align}
\frac{\partial}{\partial t}f\left(  r,t\right)  +C\frac{\partial}{\partial
r}\left(  xF\left(  x,x\right)  \right)  _{f}  &  =\overline{D}\frac{\partial
}{\partial r}\left[  \frac{\partial xF\left(  x,y\right)  }{\partial x}%
-\frac{\partial xF\left(  x,y\right)  }{\partial y}\right]  _{f}\frac
{\partial}{\partial r}f\left(  r,t\right) \nonumber\label{main}\\
&  -\frac{1}{2}C^{2}\delta t\frac{\partial}{\partial z}\left(  \frac{\partial
xF\left(  x,x\right)  }{\partial x}\right)  _{f}^{2}\frac{\partial}{\partial
r}f\left(  r,t\right)  +\mathcal{O}\left(  \epsilon^{3}\right)
\end{align}
where
\begin{align}
C  &  =\frac{\delta r}{\delta t}J_{1}\,,\nonumber\\
\overline{D}  &  =\frac{1}{2}\left(  \frac{\delta r}{\delta t}\right)
^{2}J_{2}\,.
\end{align}
\bigskip Alternatively, the diffusion coefficient can be written in terms of
the second cumulant as
\begin{equation}
D=\frac{1}{2}\frac{\left(  \delta r\right)  ^{2}}{\delta t}\left(  J_{2}%
-J_{1}^{2}\right)  ,
\end{equation}
and the equation re-arranged to give%
\begin{align}
\frac{\partial f}{\partial t}+C\frac{\partial}{\partial r}\left(  xF\left(
x,x\right)  \right)  _{f}  &  =D\frac{\partial}{\partial r}\left(
\frac{\partial xF\left(  x,y\right)  }{\partial x}-\frac{\partial xF\left(
x,y\right)  }{\partial y}\right)  _{f}\frac{\partial f}{\partial
r}\nonumber\label{main2}\\
&  +\frac{1}{2}C^{2}\delta t\frac{\partial}{\partial r}\left(  \frac{\partial
xF\left(  x,y\right)  }{\partial x}-\frac{\partial xF\left(  x,y\right)
}{\partial y}-\left(  \frac{\partial xF\left(  x,x\right)  }{\partial x}%
\right)  ^{2}\right)  _{f}\frac{\partial f}{\partial r}\,.
\end{align}
Equations (\ref{main}, \ref{main2}) give the generalized Fokker-Planck
equation and are the main result of this section.

\subsection{Effect of an external field}

If the walkers are subject to an external field, $V(r)$ , the derivation given
above must be further generalized. In stochastic algorithms, such as the
standard Metropolis Monte Carlo algorithm, the goal is to generate the
canonical distribution \cite{Frenkel}. In fact, similar reasoning also lies
behind the fluctuation-dissipation relation that is needed to specify the
autocorrelations of the noise in Langevin models. Here, we can adopt the same
approach and demand that the effect of the field be to modify the jump
probabilities so as to generate some specified steady state distribution.
Alternatively, one can adopt the position often used in modeling
nonequilibrium processes and assume that the effect of the field is the same
as in an equilibrium system - which would be equivalent to an assumption of
local equilibrium. Both possibilities are explored here.

\subsubsection{Detailed Balance}

The idea is that the stationary distribution is specified a priori as some
function of the external field. If the stationary probability to find a walker
at site $k$ is $\widehat{\pi}_{k}$ , then the master equation demands that%
\begin{equation}
0=\sum_{k=-\infty}^{\infty}\left[  \widehat{\pi}_{k}p_{l-k}F_{kl}\left(
\widehat{\pi}_{k},\widehat{\pi}_{l}\right)  -\widehat{\pi}_{l}p_{k-l}%
F_{lk}\left(  \widehat{\pi}_{l},\widehat{\pi}_{k}\right)  \right]
\end{equation}
where the subscripts on the $F$ -functions indicate that these now depend on
position via the field. The usual condition of detailed balance would be that
forward and backward jumps must balance,%
\begin{equation}
\label{balance}\widehat{\pi}_{k}p_{l-k}F_{kl}\left(  \widehat{\pi}%
_{k},\widehat{\pi}_{l}\right)  =\widehat{\pi}_{l}p_{k-l}F_{lk}\left(
\widehat{\pi}_{l},\widehat{\pi}_{k}\right)
\end{equation}
However, this assumption is problematic since the elementary probabilities
$p_{l-k}$ may make the forward and backward directions asymmetrical - in the
extreme case, backward jumps might be forbidden altogether. This is simply a
manifestation of the fact that asymmetric elementary probabilities give rise
to drift and in the case of drift it makes no sense to speak of the stationary
distribution. So we can only attempt to enforce detailed balance when the
elementary probabilities are symmetrical, in which case (\ref{balance}) reads
\begin{equation}
\widehat{\pi}_{k}F_{kl}\left(  \widehat{\pi}_{k},\widehat{\pi}_{l}\right)
=\widehat{\pi}_{l}F_{lk}\left(  \widehat{\pi}_{l},\widehat{\pi}_{k}\right)
\end{equation}
Then, making the usual separation of the jump probabilities into the
probability to generate a particular jump, $F\left(  \widehat{\pi}%
_{l-m},\widehat{\pi}_{l}\right)  $ as before, and the probability to accept a
jump, $G_{l-m,l}$ , the balance condition becomes%
\begin{equation}
\widehat{\pi}_{k}F\left(  \widehat{\pi}_{k},\widehat{\pi}_{l}\right)
G_{kl}=\widehat{\pi}_{l}F\left(  \widehat{\pi}_{l},\widehat{\pi}_{k}\right)
G_{lk}%
\end{equation}
which is solved, e.g., by the Metropolis ansatz%
\begin{equation}
G_{kl}=\min\left(  1,\frac{\widehat{\pi}_{l}F\left(  \widehat{\pi}%
_{l},\widehat{\pi}_{k}\right)  }{\widehat{\pi}_{k}F\left(  \widehat{\pi}%
_{k},\widehat{\pi}_{l}\right)  }\right)  .
\end{equation}
To proceed, we make the further assumption that the stationary distribution is
a local function of the field, $\widehat{\pi}_{l}=\Phi\left(  \beta
V_{l}\right)  =\Phi\left(  \beta V\left(  l\delta r\right)  \right)  $ . It is
shown in Appendix \ref{appfield} that in this case the generalized equation
becomes%
\begin{align}
&  \frac{\partial f}{\partial t}+\frac{\partial}{\partial r}\left(
C-D^{\prime}\left(  r\right)  K\left(  \beta V\left(  r\right)  \right)
\frac{\partial\beta V\left(  r\right)  }{\partial r} \right)  \left(
xF\left(  x,y\right)  \right)  _{f}\nonumber\label{final}\\
&  =\frac{\partial}{\partial r}\left[  \overline{D}\left(  \frac{\partial
xF\left(  x,y\right)  }{\partial x}-\frac{\partial xF\left(  x,y\right)
}{\partial y}\right)  _{f} -\frac{1}{2} C^{2}\delta t \left(  \frac{\partial
xF\left(  x,x\right)  } {\partial x}\right)  _{f}^{2} \right]  \frac{\partial
f}{\partial r}\,,
\end{align}
where%
\begin{equation}
K\left(  V\right)  =\left(  \frac{\partial\ln xF\left(  x,y\right)  }{\partial
y}-\frac{\partial\ln xF\left(  x,y\right)  }{\partial x}\right)  _{\Phi\left(
V\right)  }\frac{d}{dV}\Phi\left(  V\right)
\end{equation}
and%
\begin{equation}
D^{\prime}\left(  r\right)  =\frac{\left(  \delta r\right)  ^{2}}{\delta
t}\sum_{m=-\infty}^{\infty}m^{2}p_{m}\Theta\left(  -mK\left(  \beta V\left(
r\right)  \right)  \frac{\partial}{\partial r}\beta V\left(  r\right)
\right)  .
\end{equation}
If the elementary probabilities are symmetric, then $D^{\prime}\left(
r\right)  =D$. In this case, the advection-diffusion equation can be written
explicitly as%
\begin{align}
\frac{\partial f}{\partial t}  &  +D\frac{\partial}{\partial r}\left(
\frac{\left(  xF\left(  x,y\right)  \right)  _{f}}{\left(  xF\left(
x,y\right)  \right)  _{\Phi}}\left(  \frac{\partial xF\left(  x,y\right)
}{\partial x}-\frac{\partial xF\left(  x,y\right)  }{\partial y}\right)
_{\Phi}\frac{\partial\Phi}{\partial r}\right) \nonumber\\
&  =D\frac{\partial}{\partial r}\left(  \frac{\partial xF\left(  x,y\right)
}{\partial x}-\frac{\partial xF\left(  x,y\right)  }{\partial y}\right)
_{f}\frac{\partial f}{\partial r}\,,
\end{align}
where the fact that $f=\Phi$ is a stationary solution is obvious.

\subsubsection{Local Equilibrium and Superstatistics}

If, on the other hand, we make the local equilibrium assumption that the
acceptance probabilities are the same as in an equilibrium system,
\begin{equation}
G_{kl}=\min\left(  1,\exp\left(  -\beta\left(  V\left(  l\delta r\right)
-V\left(  k\delta r\right)  \right)  \right)  \right)
\end{equation}
then the results in Appendix \ref{appfield} give the same form as
eq.(\ref{final}), but with $K\left(  V\right)  =-1$.

The local equilibrium assumption can be relaxed by using the superstatistics
approach \cite{BeckCohen} better suited for systems out of equilibrium where
the Boltzmann distribution $\exp\left(  - \beta\left(  V \left(  r\right)
\right)  \right)  $ cannot be expected to hold. The acceptance probabilities
are then written as
\begin{equation}
\label{acc_prob}G_{k l}=\min\left(  1,\exp\left(  - \tilde\beta\left(
U\left(  l \delta r\right)  -U\left(  k\delta r\right)  \right)  \right)
\right)  \,,
\end{equation}
with
\begin{equation}
\label{superstat}\exp\left(  - \tilde\beta U( r ) \right)  = \int_{0}^{\infty}
\, d\beta\, \frac{f(\beta)}{Z(\beta)} \, e^{- \beta V(r)} \, \,,
\end{equation}
where $f(\beta)$ is a prescribed distribution of the intensive variable
$\beta$ with the normalization $Z(\beta)$. We then obtain the generalized
advection-diffusion equation (see Appendix \ref{appfield})
\begin{align}
\frac{\partial f}{\partial t}  &  +\left[  C \frac{\partial}{\partial r}-
\tilde D \left(  r\right)  \frac{\partial}{\partial r}\left(  \frac{d
\tilde\beta U\left(  r\right)  }{dr}\right)  \right]  \left(  xF\left(
x,y\right)  \right)  _{f}\nonumber\\
&  = \frac{\partial}{\partial r} \left[  D \left(  \frac{\partial xF\left(
x,y\right)  }{\partial x} - \frac{\partial xF\left(  x,y\right)  }{\partial
y}\right)  _{f} - \frac{1}{2}C^{2}\delta t \left(  \frac{\partial xF\left(
x,x\right)  }{\partial x}\right)  _{f}^{2}\right]  \frac{\partial f}{\partial
r}\,,
\end{align}
with%
\begin{equation}
\tilde D \left(  r\right)  =\frac{\left(  \delta r\right)  ^{2}}{\delta t}%
\sum_{m=-\infty}^{\infty}m^{2}p_{m}\Theta\left(  m\frac{d \tilde\beta U\left(
r\right)  }{dr}\right)  \,.
\end{equation}

\section{Properties of the Generalized Diffusion Equation}

\subsection{The Generalized Equation as a Conservation Law}

The generalized diffusion equation contains an explicit velocity, $C$.
However, since it multiplies a nonlinear function of the distribution, it is
not a drift in the usual sense that it can be eliminated by a Galilean
transformation. Although it arises from the same physical source as the drift
in classical diffusion - namely, the asymmetry of the jump probabilities - it
corresponds in the present case to a position-dependent velocity. Of course,
one could always make a transformation to an arbitrary moving frame, say with
velocity $C^{\prime}$, and this would introduce the usual term $C^{\prime
}\partial_{r}f$ into the equation.

The generalized Fokker-Planck equation can also be cast in the usual form of a
conservation law,%
\begin{equation}
\frac{\partial f}{\partial t}\,+\,\frac{\partial}{\partial r}J=0,
\end{equation}
with flux%
\begin{align}
J  &  =\left(  C-D^{\prime}\left(  r\right)  K\left(  \beta V\left(  r\right)
\right)  \frac{\partial\beta V\left(  r\right)  }{\partial r} \right)  \left(
xF\left(  x,y\right)  \right)  _{f}\nonumber\\
&  - \left[  \overline{D}\left(  \frac{\partial xF\left(  x,y\right)
}{\partial x}-\frac{\partial xF\left(  x,y\right)  }{\partial y}\right)  _{f}
\,-\,\frac{1}{2}C^{2}\delta t\left(  \frac{\partial xF\left(  x,x\right)
}{\partial x}\right)  _{f}^{2}\right]  \frac{\partial f}{\partial r}.
\end{align}
Indeed, one interpretation of the result is that it describes ordinary
diffusion with drift velocity $\mathsf{C}$ and diffusion constant $\mathsf{D}$
that are functions of the distribution, i.e.%
\begin{equation}
\frac{\partial f}{\partial t}\,+\,\frac{\partial}{\partial r}\left(
\mathsf{C}\left(  r\right)  -D^{\prime}\left(  r\right)  F\left(  f,f\right)
K\left(  \beta V\left(  r\right)  \right)  \frac{\partial\beta V\left(
r\right)  }{\partial r} \right)  f =\frac{\partial}{\partial r}\mathsf{D}%
\left(  r\right)  \frac{\partial f}{\partial r}\,,
\end{equation}
where%
\begin{align}
\mathsf{C}  &  =CF\left(  x,y\right) \nonumber\\
\mathsf{D}  &  =\overline{D}\left(  \frac{\partial xF\left(  x,y\right)
}{\partial x}-\frac{\partial xF\left(  x,y\right)  }{\partial y}\right)
_{f}-\frac{1}{2}C^{2}\delta t\left(  \frac{\partial xF\left(  x,x\right)
}{\partial x}\right)  _{f}^{2}.
\end{align}
This makes clear that in the special case $F\left(  x,y\right)  =1$, classical
diffusion is recovered.

\subsection{Scaling solutions}

We now specialize to the case that there is no drift, $C=0$, and no external
field, and ask under what circumstances a scaling solution of the form
$f\left(  r,t\right)  =t^{-\gamma/2}\phi\left(  r/t^{\gamma/2}\right)  $ is
possible:\ in other words, when does the general formulation describe
diffusion? Without drift and without external force, the generalized diffusion
equation reduces to%
\begin{equation}
\frac{\partial f}{\partial t}=\overline{D}\frac{\partial}{\partial r}\left(
M\left(  f\right)  \frac{\partial f}{\partial r}\right)
\end{equation}
where we have introduced $M(f)=\left(  \frac{\partial xF\left(  x,y\right)
}{\partial x}-\frac{\partial xF\left(  x,y\right)  }{\partial y}\right)  _{f}%
$. Defining $\zeta=r/t^{\gamma/2}$ and introducing the scaling ansatz gives%
\begin{equation}
-\frac{\gamma}{2}\left(  \phi\left(  \zeta\right)  +\zeta\frac{d}{d\zeta}%
\phi\left(  \zeta\right)  \right)  =\overline{D}t^{1-\gamma}\frac{d}{d\zeta
}\left(  M\left(  t^{-\gamma/2}\phi\left(  \zeta\right)  \right)  \frac
{d}{d\zeta}\phi\left(  \zeta\right)  \right)  \,.
\end{equation}
It is only possible to eliminate the factors of the time if $M\left(
f\right)  =m_{0}f^{\eta}$, for some constant $m_{0}$, giving
\begin{equation}
-\frac{\gamma}{2}\frac{d}{d\zeta}\zeta\phi\left(  \zeta\right)  =m_{0}%
\overline{D}t^{1-\gamma-\eta\gamma/2}\frac{d}{d\zeta}\left(  \phi^{\eta
}\left(  \zeta\right)  \frac{d}{d\zeta}\phi\left(  \zeta\right)  \right)  \,.
\end{equation}
So scaling works provided
\begin{equation}
\gamma=\frac{2}{\eta+2}\;,
\end{equation}
and the equation for the scaling function is
\begin{equation}
m_{0}\overline{D}\frac{d}{d\zeta}\left(  \phi^{\eta}\left(  \zeta\right)
\frac{d}{d\zeta}\phi\left(  \zeta\right)  \right)  +\frac{\gamma}{2}\frac
{d}{d\zeta}\zeta\phi\left(  \zeta\right)  =0\;,
\end{equation}
or%
\begin{equation}
m_{0}\overline{D}\phi^{\eta}\left(  \zeta\right)  \frac{d}{d\zeta}\phi\left(
\zeta\right)  +\frac{\gamma}{2}\zeta\phi\left(  x\right)  =A \label{phi}%
\end{equation}
for some constant, $A$. In the case $A=0$, the particular solution is easily
found from
%\begin{align}
$\frac{d}{d\zeta}\phi^{\eta}\left(  \zeta\right)  =-\frac{\eta\gamma}%
{2m_{0}\overline{D}}\,\zeta$, and is given by
\begin{equation}
\phi\left(  \zeta\right)  =\left(  B-\frac{\eta}{2\left(  2+\eta\right)
m_{0}\overline{D}}\zeta^{2}\right)  ^{1/\eta}\Theta\left(  B-\frac{\eta
}{2\left(  2+\eta\right)  m_{0}\overline{D}}\zeta^{2}\right)  \,.
\end{equation}
The constant $B$ is determined by normalization:%
\begin{equation}
B=\left(  \frac{\eta\left(  \eta+2\right)  }{8m_{0}\overline{D}}\right)
^{\frac{\eta}{\eta+2}}\allowbreak B\left(  \frac{1}{\eta},\frac{1}{2}\right)
^{-\frac{2\eta}{\eta+2}}%
\end{equation}
where $B(x,y)=\Gamma\left(  x\right)  \Gamma\left(  y\right)  /\Gamma\left(
x+y\right)  $ is the beta function. The distribution can also be written as a
$q$-Gaussian by defining $\eta=1-q$,%
\begin{equation}
\phi\left(  \zeta\right)  =B^{\frac{1}{1-q}}\left(  1-\frac{1-q}%
{2Bm_{0}\overline{D}\left(  3-q\right)  }\zeta^{2}\right)  ^{\frac{1}{1-q}%
}\Theta\left(  1-\frac{1-q}{2Bm_{0}\overline{D}\left(  3-q\right)  }\zeta
^{2}\right)  .
\end{equation}

Note that the scaling hypothesis demands that%
\begin{equation}
M(f)\equiv\left(  \frac{\partial xF\left(  x,y\right)  }{\partial x}%
-\frac{\partial xF\left(  x,y\right)  }{\partial y}\right)  _{f}=m_{0}f^{\eta
}\,,
\end{equation}
and the fact that the function $F$ is defined in terms of the jump
probabilities means that it must be bounded. From its definition, we expect
that
\begin{equation}
0\leq xF\left(  x,y\right)  \leq1\;\;\;\;\mbox{and}\;\;\;\;0\leq yF\left(
x,y\right)  \leq1\,,
\end{equation}
for all $x,y\in\left[  0,1\right]  $. If for example $F(x,y)=F(x)$, then the
scaling hypothesis is: $F(x)\sim x^{\eta}$, so that the bounds given above
demand that
\begin{equation}
\eta\geq0\,,\;\;\;\;0\leq\gamma\leq1\,,\;\;\;\;\mbox{and}\;\;\;\;q\leq1\ \,.
\label{limits}%
\end{equation}

All of the preceding concerning the scaling behavior only applies to the case
that the constant $A$ is taken to be zero in eq. (\ref{phi}). For values of
$A\neq0$, no general solution of this equation could be found. However, note
that if $\phi\left(  \zeta\right)  $ is analytic at $\zeta=0$, then from
eq.(\ref{phi}) 
\begin{equation}
\lim_{\zeta\rightarrow0}\frac{d}{d\zeta}\phi\left(  \zeta\right)  =\frac
{A}{m_{0}\overline{D}\phi^{\eta}\left(  0\right)  }%
\end{equation}
so that any solution with $A\neq0$ is not symmetric about $\zeta=0$. Thus, we
can say that the scaling behavior discussed here applies to the general case
of symmetric solutions.

\section{Numerical Tests}

In order to test the validity of the generalized diffusion equation, we have
performed numerical simulations of the underlying random walk model. Our
simulations begin with a population of $N$, independent random walkers at
position $r=0$ at time $t=0$. At time step $i$, each walker makes a jump of
$m$ lattice sites from its present position, say site $l$, with a probability:
$p_{m}F\left(  \widehat{f}_{l}\left(  i\right)  ,\widehat{f}_{l+m}\left(
i\right)  \right)  $ where the distribution $\widehat{f}_{k}\left(  i\right)
$ is simply the fraction of walkers at site $k$ at time step $i$. All of the
simulations discussed below were performed using a population of size
$N=10^{5}$.

\begin{figure}[ptb]
\begin{center}
\resizebox{12cm}{!}{
{\includegraphics[height = 12cm, angle=-90]{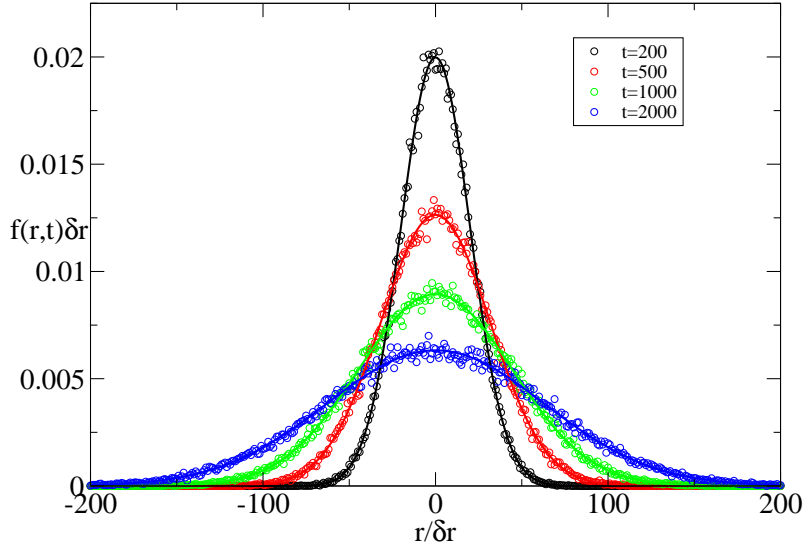}}}
\end{center}
\caption{(Color online) The evolution of an initial delta-function distribution for the case
$q=0.999$ and equal elementary probabilities for jumps up to length 2. The
symbols are from Monte Carlo simulation of the random-walk and the solid lines
are the analytic $q$-Gaussian solution (\ref{qg}) to the generalized diffusion
equation.}%
\label{fig1}%
\end{figure}

\begin{figure}[tbp]
\begin{center}
\resizebox{12cm}{!}{
{\includegraphics[height = 12cm, angle=-90]{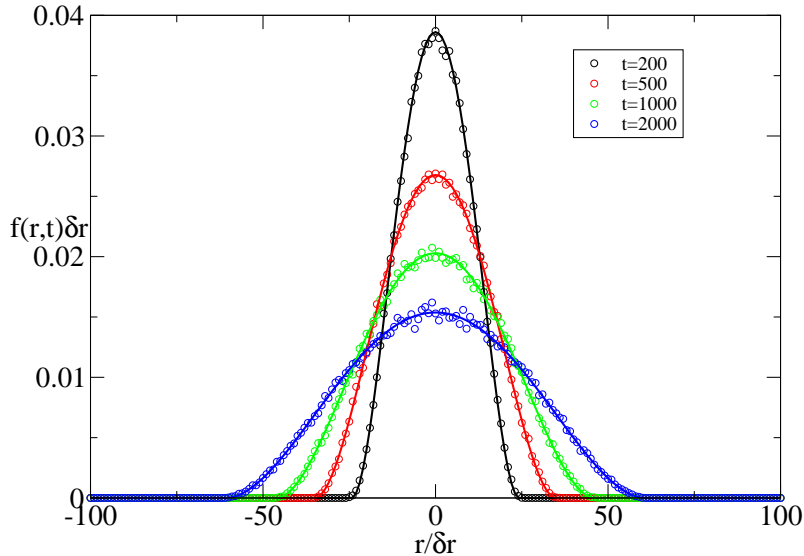}}}
\end{center}
\caption{(Color online) Same as Fig. \ref{fig1}, but for $q=0.5$.}%
\label{fig2}%
\end{figure}

\begin{figure}[tbp]
\begin{center}
\resizebox{12cm}{!}{
{\includegraphics[height = 12cm, angle=-90]{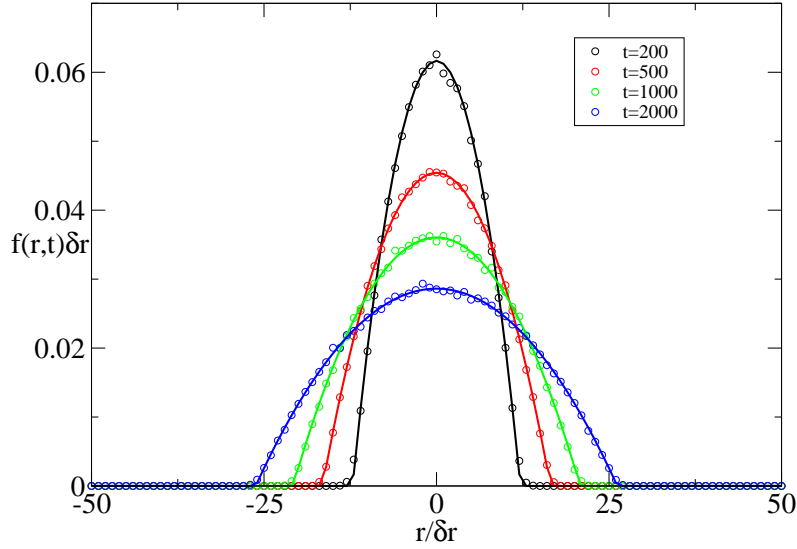}}}
\end{center}
\caption{(Color online) Same as Fig. \ref{fig1}, but for $q=0.0$.}%
\label{fig3}%
\end{figure}

The first simulation is for the case of no drift, elementary probabilities
$p_{j}=\frac{1}{5}$ with $j\in\left[  -2,2\right]  $, and $F(x,y)=x^{1-q}$ for
which the theory gives%
\begin{align}
f\left(  r,t\right)   &  =t^{-\frac{1}{3-q}}\phi_{q}\left(  r/t^{\frac{1}%
{3-q}}\right) \nonumber\label{qg}\\
\phi_{q}\left(  \zeta\right)   &  =B_{q}^{\frac{1}{1-q}}\left(  1-\frac
{1-q}{2B_{q}\left(  2-q\right)  \left(  3-q\right)  \overline{D}}\,\zeta
^{2}\right)  ^{\frac{1}{1-q}}\Theta\left(  1-\frac{1-q}{2B_{q}\left(
2-q\right)  \left(  3-q\right)  \overline{D}}\,\zeta^{2}\right) \nonumber\\
B_{q}  &  =\left(  \frac{\left(  1-q\right)  \left(  3-q\right)  }{8\left(
2-q\right)  \overline{D}}\right)  ^{\frac{1-q}{3-q}}\allowbreak B\left(
\frac{1}{1-q},\frac{1}{2}\right)  ^{-\frac{2-2q}{3-q}}\,.
\end{align}
As stated above, only the range $q \leq1$ is permitted, and the value $q=1$
corresponds to classical diffusion. Since the initial condition and jump
probabilities are symmetric, there is no drift and the scaling solution
applies. Figures (\ref{fig1}-\ref{fig3}) show the analytic results, given by
eq.(\ref{qg}), and the results of the microscopic simulations for $q=0.999$
(essentially the classical case), $q=0.5$ and $q=0.0$. These correspond to
anomalous diffusion with scaling exponent $\gamma=0.999\,5$, $\frac{4}{5}$ and
$\frac{2}{3}$ respectively. In all cases, the agreement between the
simulations and the scaling solution is very good, even at the earliest times.

\begin{figure}[ptb]
\begin{center}
\resizebox{12cm}{!}{
{\includegraphics[height = 12cm, angle=-90]{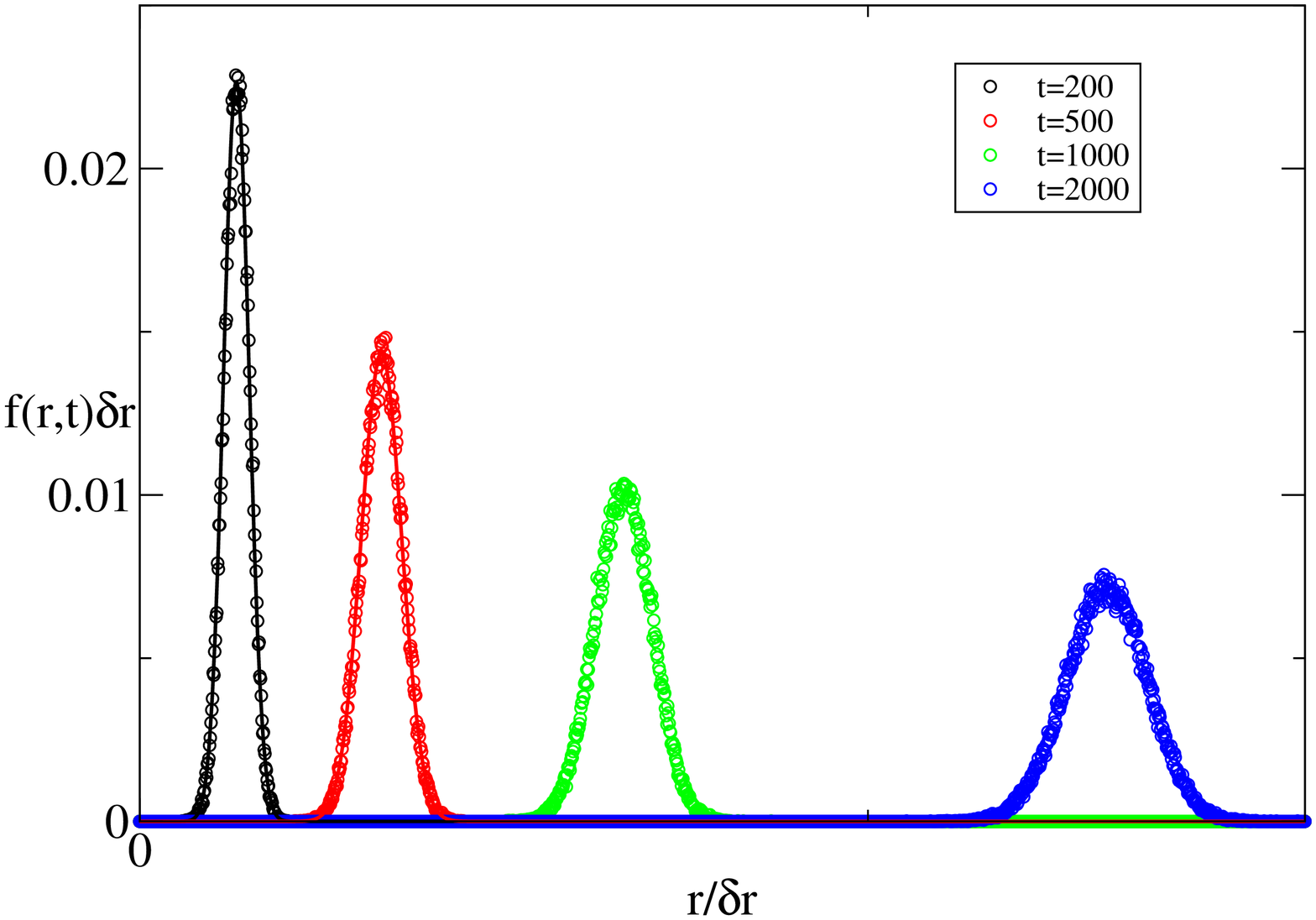}}}
\end{center}
\caption{(Color online) The evolution of an initial delta-function distribution for the case
$q=0.999$ and $p_{j}=\frac{j+3}{15}$ for $j\in\left[  -2,2\right]  $. Since
the probabilities violate detailed balance, there is a non-zero drift
velocity, $C = \frac{2\delta r}{3\delta t}$. The symbols are from Monte Carlo
simulation of the random-walk and the solid lines are the numeric solution of
the generalized diffusion equation, eq.(\ref{main2}).}%
\label{fig4}%
\end{figure}

\begin{figure}[tbp]
\begin{center}
\resizebox{12cm}{!}{
{\includegraphics[height = 12cm, angle=-90]{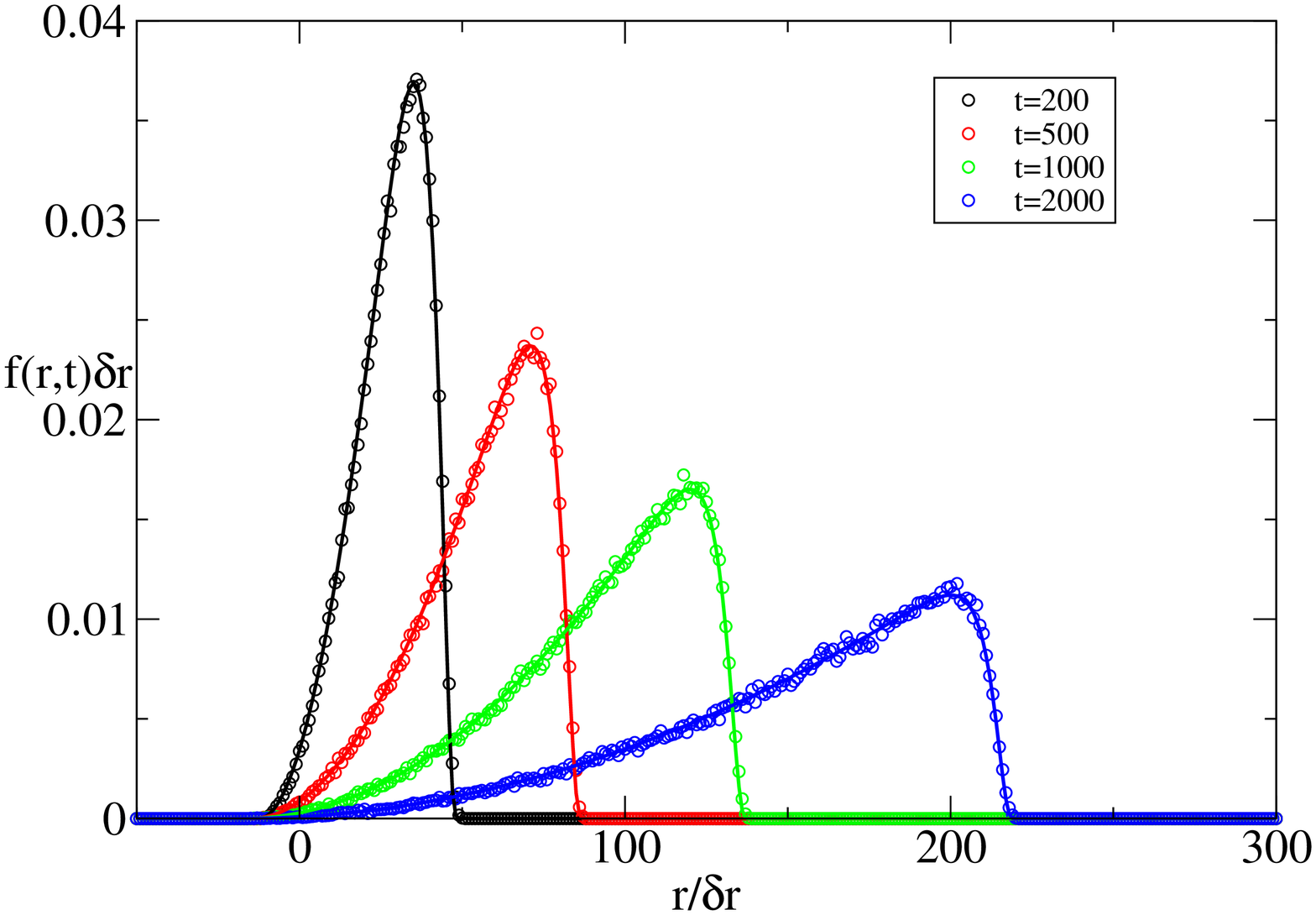}}}
\end{center}
\caption{(Color online) Same as Fig. \ref{fig4}, but for $q=0.5$.}%
\label{fig5}%
\end{figure}

\begin{figure}[ptb]
\begin{center}
\resizebox{12cm}{!}{
{\includegraphics[height = 12cm, angle=-90]{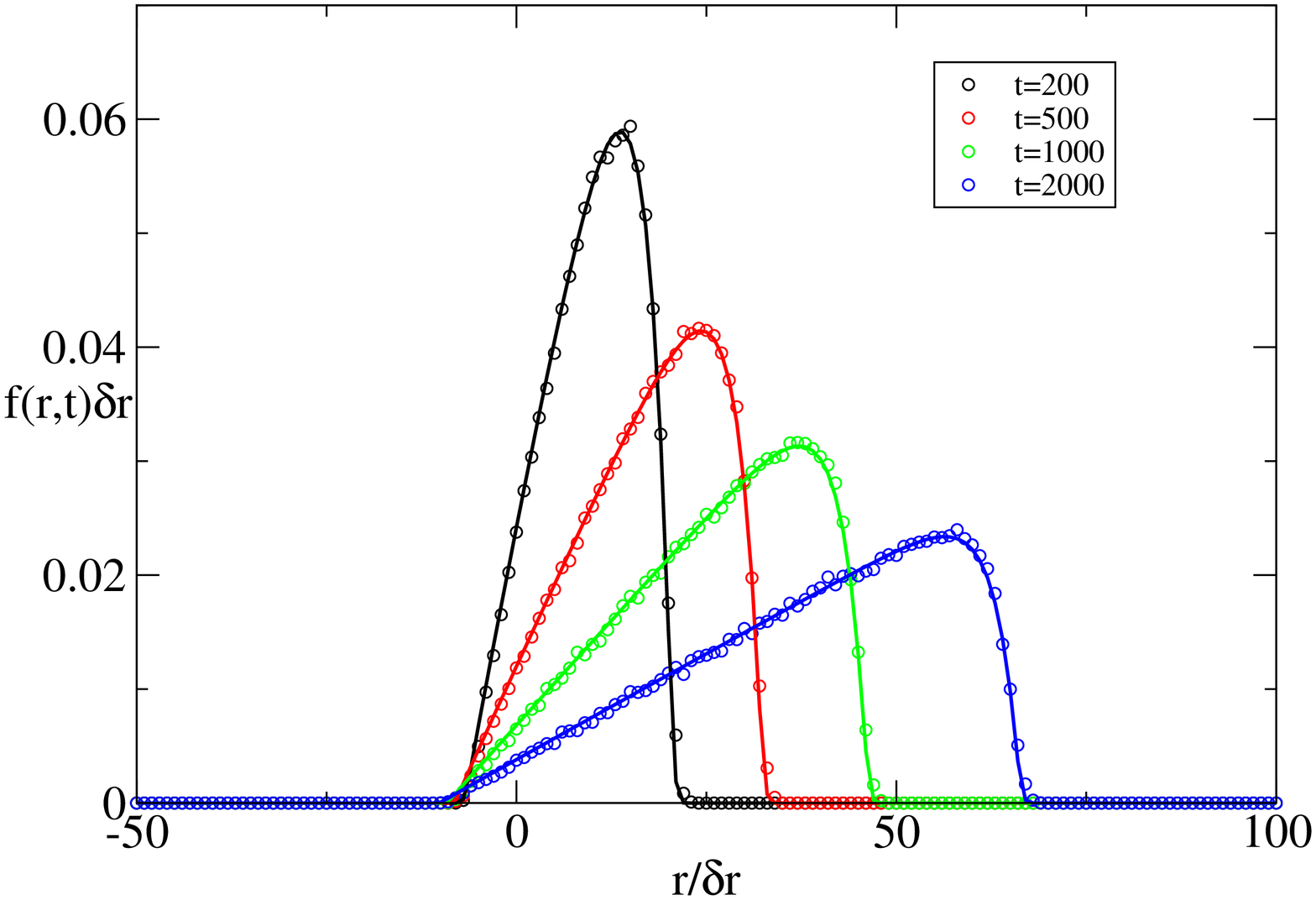}}}
\end{center}
\caption{(Color online) Same as Fig. \ref{fig4}, but for $q=0.0$.}%
\label{fig6}%
\end{figure}

In the second set of simulations, particles are subjected to drift. In this
case, the elementary probabilities are taken to be $p_{j}=\frac{j+3}{15}$ for
$j\in\left[  -2,2\right]  $. Figures (\ref{fig4}-\ref{fig6}) show the
evolution of the distributions for the same values of $q$ as for the no-drift
case. As mentioned above, it is then no longer possible to solve the
generalized diffusion equation analytically. So comparison is made to a
numerical solution of eq.(\ref{main2}) with $F(x,y)=x^{1-q}$. The numerical
solution was performed using centered finite differences in the spatial
variable and a simple, first-order scheme in the time, with the lattice
spacing fixed at $\delta r$ and the time step equal to $0.001\delta t$. For
$q=0.999$, the process is essentially that of the classical case of
advection-diffusion. For the smaller values of $q$ however, the distribution
is very different, becoming increasingly asymmetrical as time progresses. As
$q$ becomes smaller, and the processes becomes more sub-diffusive, the
velocity of the peak of the distribution also decreases. Even with these
significant qualitative changes for decreasing values of $q$, the generalized
diffusion equation is again seen to give very good agreement with the Monte
Carlo simulations.

\begin{figure}[h!tb]
\begin{center}
\resizebox{12cm}{!}{
{\includegraphics[height = 12cm, angle=-90]{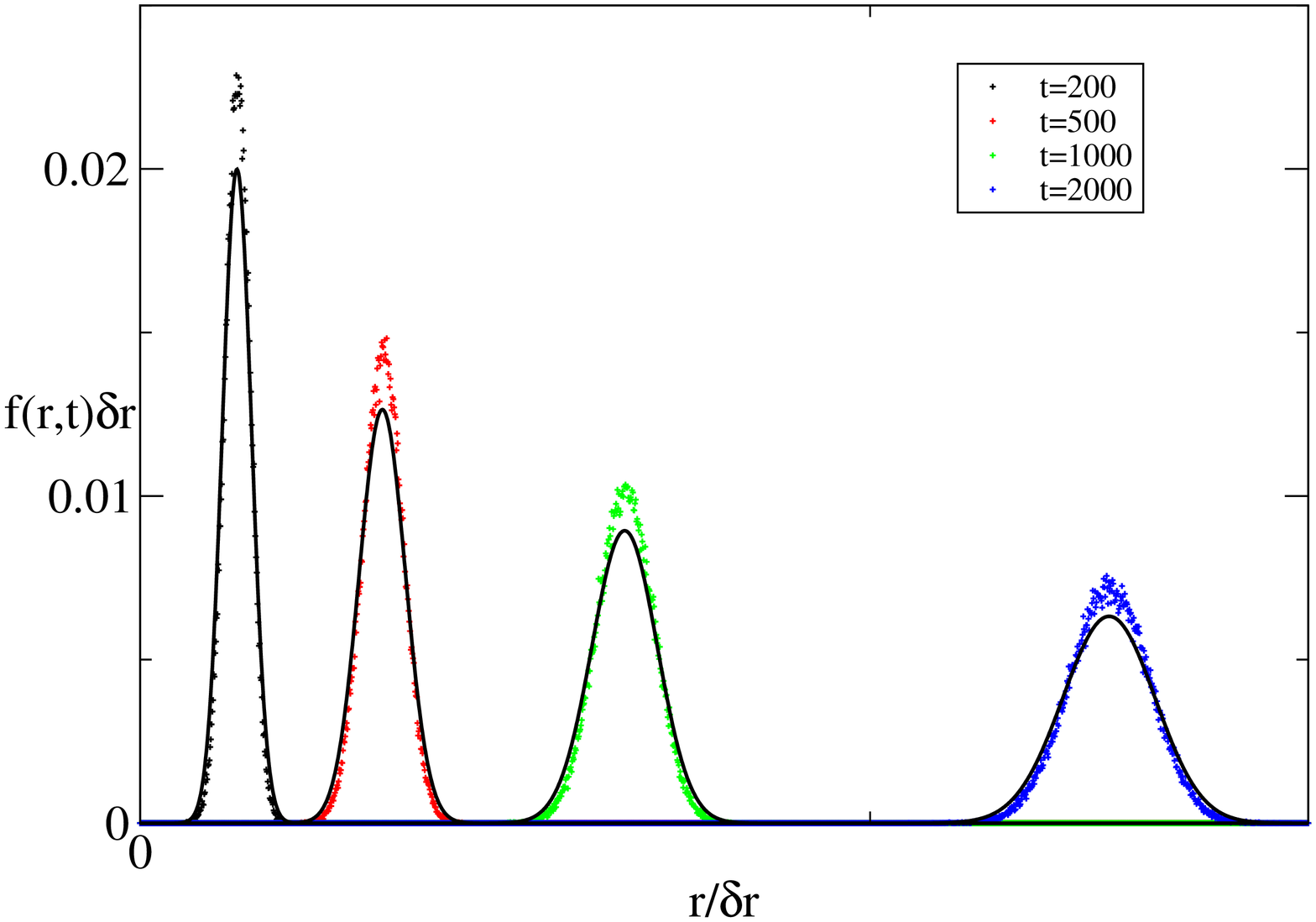}}}
\end{center}
\caption{(Color online) Simulation data for $q=0.999$ and the predictions of
the Fokker-Planck equation with the type I modification. The type II is not
shown as it gives virtually the same result as the full Fokker-Planck
equation, as shown in Fig. \ref{fig4}, and is in almost perfect agreement with
the data.}%
\label{fig7}%
\end{figure}

\begin{figure}[ptb]
\begin{center}
\resizebox{12cm}{!}{
{\includegraphics[height = 12cm, angle=-90]{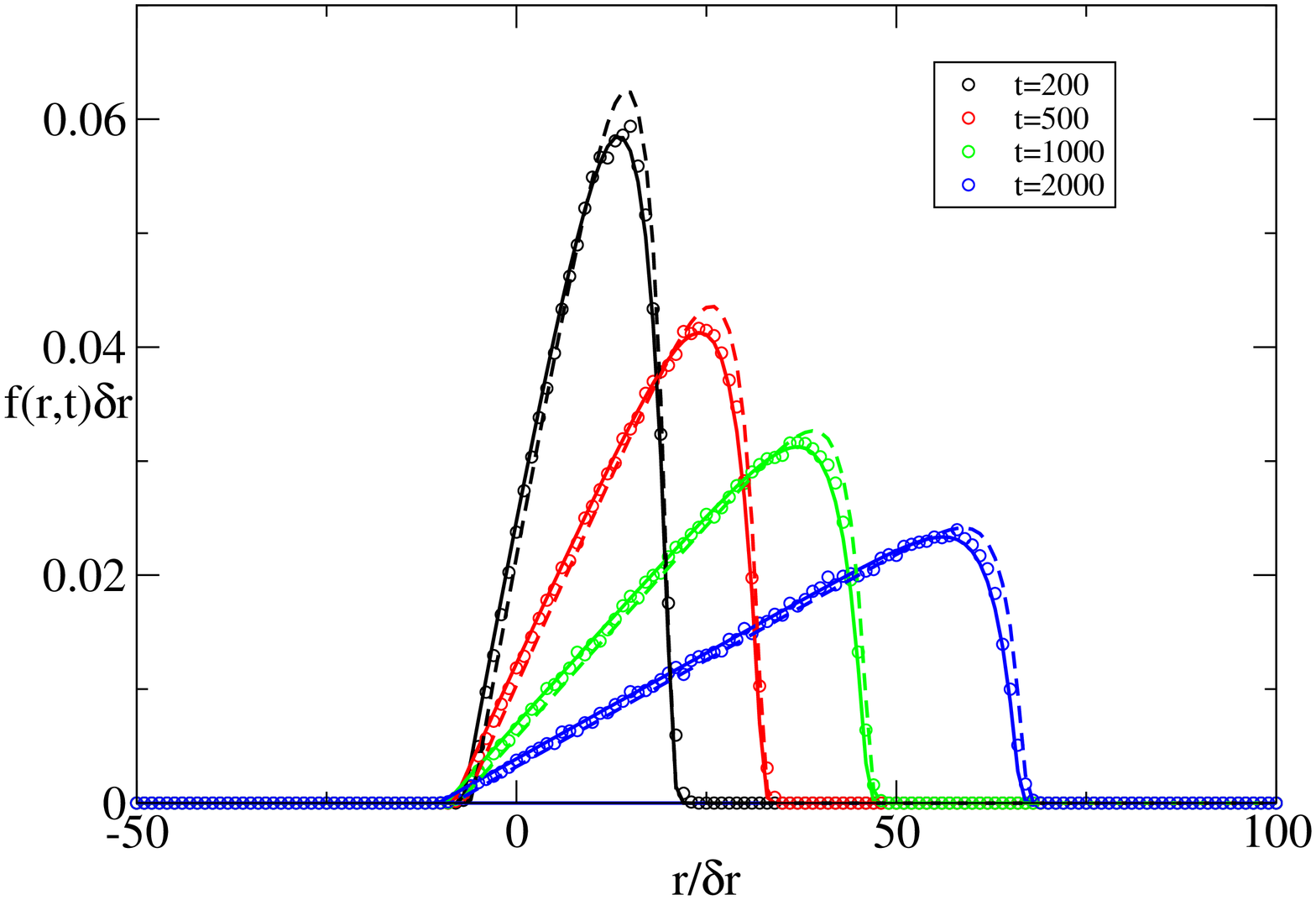}}}
\end{center}
\caption{(Color online) The simulation data for $q=0$ together with the predictions of the
Fokker-Planck equation with the type I (full line) and type II (broken line)
modifications.}%
\label{fig8}%
\end{figure}

One interesting question is whether the new terms appearing in the generalized
diffusion equation (\ref{main}, \ref{main2}) play any role, or whether they
could be neglected giving a result closer to the Porous Media Equation
\cite{muskat} which (with drift term) reads
\begin{equation}
\frac{\partial}{\partial t}f\left(  r,t\right)  + C \frac{\partial}{\partial
t}f^{\alpha}\left(  r,t\right)  = D\frac{\partial^{2}}{\partial r^{2}%
}f^{\alpha}\left(  r,t\right)  \,. \label{PME}%
\end{equation}
%with $\alpha + q = 2$.
To investigate this, we repeated the solution of two modifications of the
generalized diffusion equation. In the first case, the "extra" terms are
simply omitted from eq.(\ref{main}) giving
\begin{equation}
\frac{\partial f}{\partial t} +C\frac{\partial}{\partial r}\left(  xF\left(
x,y\right)  \right)  _{f}=\overline{D}\frac{\partial}{\partial r}\left(
\frac{\partial xF\left(  x,y\right)  }{\partial x}-\frac{\partial xF\left(
x,y\right)  }{\partial y}\right)  _{f}\frac{\partial f}{\partial r}\,.
\label{modI}%
\end{equation}
One objection to this approximation is that it does not reduce to the expected
result in the limit of classical diffusion, since then the extra term would
combine with the diffusive term to make the replacement $\overline
{D}\rightarrow D$. This leads to the second modification considered here,
%namely the same as the first case but with the alternative form of the generalized
namely omitting the extra term from equation~(\ref{main2}) which then reads
\begin{equation}
\frac{\partial f}{\partial t} +C\frac{\partial}{\partial r}\left(  xF\left(
x,y\right)  \right)  _{f}={D}\frac{\partial}{\partial r}\left(  \frac{\partial
xF\left(  x,y\right)  }{\partial x}-\frac{\partial xF\left(  x,y\right)
}{\partial y}\right)  _{f}\frac{\partial f}{\partial r}\,. \label{modIl}%
\end{equation}
For want of better terms, these will be referred to as modifications I and
II\ respectively. Figures (\ref{fig7}) and (\ref{fig8}) show the numerical
solution of these equations compared to the simulation data for the two cases
$q=0.999$ and $q=0$. For $q=0.999$, the type II modification is a much better
approximation to the data than is the type I modification as might be expected
since type II becomes exact for $q=1$. However, the results for $q=0$ are
exactly reversed: type I is a noticeably better approximation than is type II.
The conclusion is that the full equation is necessary to provide a good
description of the system for all values of $q$. 

\section{Conclusions}

We have shown that a simple modification of the classical random walk gives
rise to sub-diffusive behavior. The required modification is that the
probability to make a jump from one lattice site to another depends on the
occupation probability of the walker on the lattice. Using a multiscale
expansion of the exact master equation, we derived a generalized Fokker-Planck
equation. In the limit of symmetric probabilities to jump left and right, this
equation gives rise to diffusive behavior of the moments of the distribution
provided the dependence of the jump probabilities takes the form of a power
law. In this case, our result reduces to the well known Porous Media equation.
Unlike other approaches that begin with a continuous time random walk, we
specifically consider a microscopic model in the \emph{hydrodynamic} limit of
large length and time scales. This is responsible for the appearance of a new
term in the generalized diffusion equation which, as comparison to simulations
of the microscopic model shows, is necessary to correctly describe the
evolution of the distribution.

Our generalized equation reduces to previous results in the appropriate
limits. Most simply, if the function $F(x,y)=1$ the Fokker-Planck equation
becomes the classical advective-diffusive equation. The continuous time random
walk results from the scaling%
\begin{equation}
\delta t\rightarrow\epsilon^{2}\delta t\;,\;\;\;\;\delta r\rightarrow
\epsilon\delta r\;,\;\;\;\;J_{1}\rightarrow\epsilon J_{1}\;,
\end{equation}
and the limit $\epsilon\rightarrow0$. (Note that this limit is easily deduced
directly from the smoothed master equation without need for the multiscale
expansion.) With the further approximations of (i) no drift ( $C=0$) and (ii)
hops of only one lattice site our result agrees with that of Curado and Nobre
\cite{CuradoNobre}.

We have shown that exact self-similar solutions of the generalized diffusion
equation (without drift) are only possible if the jump probabilities scale as
power laws. In this case, the distribution turns out to be the so-called
$q$-Gaussian often introduced in an ad hoc manner to describe anomolous
diffusion. The model presented here therefore gives one answer to the question
"what underlying dynamics could give rise to the observed $q$-Gaussian
distributions":\ a dependence of the jump probabilities on the local
distribution (or more likely, local concentration) of walkers is sufficient.
Note that the dependence need not be an exact power-law:\ it is enough that
the long-time limit of the diffusion equation admits scaling which in turn
implies that the function $F\left(  x,y\right)  $ become algebraic in $x$ in
the limit of very small or very large $x$ depending on the various scaling
exponents. One restriction of the exact scaling result, however, is that our
model is only well defined if $F(x,y)=x^{\eta}$ for $\eta>0$ which in turn
implies sub-diffusive scaling of the moments. To describe super-diffusion
there are only two possibilities. Either one could construct a function
$F\left(  x,y\right)  $ that gives the proper normalization of the jump
probabilities and that gives super-diffusion in the long-time limit or one
could modify the basic description of the jump probabilities, eq.(\ref{jumps})
so as to introduce nonlinearity in some other way.

The generalized diffusion equation (GDE) is in some ways similar to the
fractional Fokker-Planck equation (FFPE): both describe sub-diffusion and both
require power-law probabilities to give the subdiffusion (the GDE in the jump
probabilities, the FFPE in the waiting times). It is natural to ask whether,
given some experimental data which shows subdiffusion, there is any way to
choose between the two descriptions. On physical grounds, the idea of waiting
times that are distributed as a power law might be more appropriate in which
case the FFPE should be prefered; if there it makes more sense to think in
terms of an interaction between the walkers, then the GDE\ might be more
appropriate. Empirically, if the distribution of walkers is measured, it might
be possible to choose a model based on the fact that the GDE predicts that the
distribution of walkers in a system showing subdiffusion with no external
forces should obey a $q$-Gaussian distribution whereas in the case of the FFPE
there is also a scaling solution but the distribution is a stretched
gaussian\cite{Metzler}. In fact, in two studies, one of
sub-diffusion induced by a random walk on a Sierpinski
gasket\cite{Schulsky} and the other of super-diffuson induced by a
raondom walk on a tree structure\cite{Fischer}, it was shown that the
FFPE and GDE results were sufficiently different as to allow an
empirical distinction to be made.  

In the presence of either drift (ie non-symmetric jump probabilities) or an
external field, the GDE\ is more complex than the equilvalent extension of the
Porous Media equation. This is true even when a power-law dependence of the
jump probabilities is assumed since in this case, the GDE\ becomes%
\begin{align}
&  \frac{\partial f}{\partial t}+\frac{\partial}{\partial r}\left(  C+\left(
1+\eta\right)  D^{\prime}\left(  r\right)  \frac{\partial\ln\Phi\left(  \beta
V\left(  r\right)  \right)  }{\partial r}\right)  f^{1+\eta}\nonumber\\
&  =\frac{\partial}{\partial r}\left[  \left(  1+\eta\right)  \overline
{D}f^{\eta}-\frac{1}{2}\left(  1+\eta\right)  ^{2}C^{2}\delta tf^{2\eta
}\right]  \frac{\partial f}{\partial r}\,,
\end{align}
where%
\begin{equation}
D^{\prime}\left(  r\right)  =\frac{\left(  \delta r\right)  ^{2}}{\delta
t}\sum_{m=-\infty}^{\infty}m^{2}p_{m}\Theta\left(  \left(  1+\eta\right)
m\frac{\partial\ln\Phi\left(  \beta V\left(  r\right)  \right)  }{\partial
r}\right)  .
\end{equation}
or, with $1 + \eta = \alpha$,
\begin{align}
 \frac{\partial f}{\partial t}+\frac{\partial}{\partial r}\left(  C+
\alpha  D^{\prime}\left(  r\right)  \frac{\partial\ln\Phi\left(  \beta
V\left(  r\right)  \right)  }{\partial r}\right)  f^{\alpha}
  =\frac{\partial}{\partial r}\left[{D} + \frac{\delta t}{2}C^{2}\left(1-\alpha f^{\alpha -1}\right) \right]  \frac{\partial }{\partial r}f^{\alpha}\,,
\end{align}
to be compared with the PME, eq.(\ref{PME}).
With no field, the drift does not generate a simple Galilean transformation of
the equation without drift, as is usually assumed to be the case with the
Porous Media equation, but instead generates new nonlinearities in the GDE.
Because the drift term has the same number of powers of $f$ but one fewer
derivative, than the right hand side, no simple scaling solution is evident.
In the case of an external field but no drift, one has that $D^{\prime}\left(
r\right)  \rightarrow\overline{D}$ so that the gradient terms on the left and
right hand sides of the equation have the same number of powers of $f$ and of
spatial gradients. A scaling solution would then be possible, but only with a
trivial external field. This superficial analysis suggests that exact scaling
is only possible in the GDE in the case of no field and no drift. It leaves
open the possibility of approximate scaling in the long-time limit, not to
mention the possibility that more complex assumptions for the dependence of
the jump-probabilities might give completely different scaling properties.
These questions are the subject of on-going research. 

\bigskip

\bigskip\acknowledgments The work of JFL was supported in part by the European
Space Agency under contract number ESA AO-2004-070. The authors thank
Patrick Grosfils for several useful dicussions.

\bigskip

\appendix

\section{Role of the external field}

\label{appfield}

The master equation is
\begin{equation}
f\left(  r,t+\delta t\right)  =f\left(  r,t\right)  +\sum_{m=-\infty}^{\infty
}p_{m}\left[
\begin{array}
[c]{c}%
f\left(  r-m\delta r,t\right)  F\left(  f\left(  r-m\delta r,t\right)
,f\left(  r,t\right)  \right)  G_{r-m,r}\\
-f\left(  r,t\right)  F\left(  f\left(  r,t\right)  ,f\left(  r+m\delta
r,t\right)  \right)  G_{r,r+m}%
\end{array}
\right]
\end{equation}
where%
\begin{equation}
G_{l-m,l}=\frac{\Phi_{l}F\left(  \Phi_{l},\Phi_{l-m}\right)  }{\Phi
_{l-m}F\left(  \Phi_{l-m},\Phi_{l}\right)  }\Theta\left(  1-\frac{\Phi
_{l}F\left(  \Phi_{l},\Phi_{l-m}\right)  }{\Phi_{l-m}F\left(  \Phi_{l-m}%
,\Phi_{l}\right)  }\right)  +\Theta\left(  \frac{\Phi_{l}F\left(  \Phi
_{l},\Phi_{l-m}\right)  }{\Phi_{l-m}F\left(  \Phi_{l-m},\Phi_{l}\right)
}-1\right)
\end{equation}
and%
\begin{equation}
\Phi_{l}=\Phi\left(  V\left(  l\delta r\right)  \right)  .
\end{equation}
This can also be written as
\begin{align}
G_{l-m,l}  &  =1+\left(  \frac{\Phi_{l}F\left(  \Phi_{l},\Phi_{l-m}\right)
}{\Phi_{l-m}F\left(  \Phi_{l-m},\Phi_{l}\right)  }-1\right)  \Theta\left(
1-\frac{\Phi_{l}F\left(  \Phi_{l},\Phi_{l-m}\right)  }{\Phi_{l-m}F\left(
\Phi_{l-m},\Phi_{l}\right)  }\right) \\
&  =1+\left(  \frac{\Phi_{l}F\left(  \Phi_{l},\Phi_{l-m}\right)  }{\Phi
_{l-m}F\left(  \Phi_{l-m},\Phi_{l}\right)  }-1\right)  \Theta\left(
\Phi_{l-m}F\left(  \Phi_{l-m},\Phi_{l}\right)  -\Phi_{l}F\left(  \Phi_{l}%
,\Phi_{l-m}\right)  \right) \nonumber\\
&  =1+H_{l-m,l}\nonumber
\end{align}
where%
\begin{equation}
H_{l-m,l}=\left(  \frac{h\left(  l\delta r,\left(  l-m\right)  \delta
r\right)  }{h\left(  \left(  l-m\right)  \delta r,l\delta r\right)
}-1\right)  \Theta\left(  1-\frac{h\left(  l\delta r,\left(  l-m\right)
\delta r\right)  }{h\left(  \left(  l-m\right)  \delta r,l\delta r\right)
}\right) \nonumber
\end{equation}
with%
\begin{equation}
h\left(  x,y\right)  =\Phi\left(  V\left(  x\right)  \right)  F\left(
\Phi\left(  V\left(  x\right)  \right)  ,\Phi\left(  V\left(  y\right)
\right)  \right)  .
\end{equation}
The goal is to develop the expansion of $H_{l-m,l}$ in terms of $\delta r$ and
to use this to derive the modified advection-diffusion equation. In the
Appendix, we use an abbreviated notation whereby $\partial_{r}=\frac{\partial
}{\partial r}$, etc.

\bigskip First, note that for present purposes we need the expansion of
$H_{l-m,l}$ up to order $\left(  \delta r\right)  ^{2}$ inclusive. For the
step-function part, we have
\begin{align}
\Theta\left(  1-\frac{h\left(  x,x-u\right)  }{h\left(  x-u,x\right)
}\right)   &  =\Theta\left(  h\left(  x-u,x\right)  -h\left(  x,x-u\right)
\right) \nonumber\\
&  =\Theta\left(  u\left(  h_{y}-h_{x}\right)  +\frac{1}{2}u^{2}\left(
h_{xx}-h_{yy}\right)  +...\right) \nonumber\\
&  =\Theta\left(  \frac{u}{\left\vert u\right\vert }\left(  h_{y}%
-h_{x}\right)  +\frac{1}{2}\left\vert u\right\vert \left(  h_{xx}%
-h_{yy}\right)  +...\right)
\end{align}
Now, we assume that $\left(  h_{y}-h_{x}\right)  $ is of order one, so that in
some formal sense we can expand to get%
\begin{equation}
\Theta\left(  1-\frac{h\left(  x,x-u\right)  }{h\left(  x-u,x\right)
}\right)  =\Theta\left(  \frac{u}{\left\vert u\right\vert }\left(  h_{y}%
-h_{x}\right)  \right)  +\frac{1}{2}\left\vert u\right\vert \left(
h_{xx}-h_{yy}\right)  \delta\left(  \frac{u}{\left\vert u\right\vert }\left(
h_{y}-h_{x}\right)  \right)  +...
\end{equation}
In general, the $\delta$-function (and higher order terms) only contribute on
a set of measure zero and can be neglected. (Furthermore, we will explicitly
show that the delta function cannot contribute until at least order $\left(
\delta r\right)  ^{3}$.) Expanding the coefficient of the step function gives%
\begin{equation}
\left(  \frac{h\left(  x,x-u\right)  }{h\left(  x-u,x\right)  }-1\right)
=u\frac{h_{x}-h_{y}}{h}+\frac{1}{2}u^{2}\left(  \frac{h_{yy}}{h}-2\frac
{h_{x}h_{y}}{h^{2}}+2\frac{h_{x}^{2}}{h^{2}}-\frac{h_{xx}}{h}\right)  +...
\end{equation}
Multiplying these two contributions, we see that the $\delta$-function first
appears at order $u^{2}$, but in the form of $x\delta\left(  x\right)  $ which
is always zero; so, as stated above, it cannot contribute until at least order
$u^{3}$, if at all. The result is
\begin{align}
&  \left(  \frac{h\left(  x,x-u\right)  }{h\left(  x-u,x\right)  }-1\right)
\Theta\left(  1-\frac{h\left(  x,x-u\right)  }{h\left(  x-u,x\right)  }\right)
\nonumber\\
&  =\left[  u\left(  \frac{h_{x}-h_{y}}{h}\right)  +\frac{1}{2}u^{2}\left(
\frac{h_{yy}}{h}-2\frac{h_{x}h_{y}}{h^{2}}+2\frac{h_{x}^{2}}{h^{2}}%
-\frac{h_{xx}}{h}\right)  \right]  \Theta\left(  u\left(  h_{y}-h_{x}\right)
\right)  +\mathcal{O}\left(  u^{3}\right)  \,,
\end{align}
and consequently%
\begin{equation}
H_{l-m,l}=\left[  \left(  m\delta r\right)  \left(  \frac{h_{x}-h_{y}}%
{h}\right)  +\frac{1}{2}\left(  m\delta r\right)  ^{2}\left(  \frac{h_{yy}}%
{h}-2\frac{h_{x}h_{y}}{h^{2}}+2\frac{h_{x}^{2}}{h^{2}}-\frac{h_{xx}}%
{h}\right)  \right]  \Theta\left(  m\left(  h_{y}-h_{x}\right)  \right)  \,.
\end{equation}
Similarly%
\begin{align}
H_{l,l+m}  &  =\left(  \frac{\Phi_{l+m}F\left(  \Phi_{l+m},\Phi_{l}\right)
}{\Phi_{l}F\left(  \Phi_{l},\Phi_{l+m}\right)  }-1\right)  \Theta\left(
\Phi_{l}F\left(  \Phi_{l},\Phi_{l+m}\right)  -\Phi_{l+m}F\left(  \Phi
_{l+m},\Phi_{l}\right)  \right) \nonumber\\
&  =\left(  \frac{h\left(  \left(  l+m\right)  \delta r,l\delta r\right)
}{h\left(  l\delta r,\left(  l+m\right)  \delta r\right)  }-1\right)
\Theta\left(  1-\frac{h\left(  \left(  l+m\right)  \delta r,l\delta r\right)
}{h\left(  l\delta r,\left(  l+m\right)  \delta r\right)  }\right) \nonumber\\
&  =\left(  \left(  m\delta r\right)  \frac{h_{x}-h_{y}}{h}+\frac{1}{2}\left(
m\delta r\right)  ^{2}\left(  \frac{h_{xx}}{h}-\frac{2h_{x}h_{y}}{h^{2}}%
+\frac{2h_{y}^{2}}{h^{2}}-\frac{h_{yy}}{h}\right)  \right)  \Theta\left(
\left(  m\delta r\right)  \left(  h_{y}-h_{x}\right)  \right) \nonumber\\
&  + \mathcal{O} \left(  m^{3}\right)
\end{align}
Substituting back into the master equation gives%
\begin{align}
&  f\left(  r,t+\delta t\right)  =f\left(  r,t\right) \nonumber\\
&  +\sum_{m=-\infty}^{\infty}p_{m}\left[  f\left(  r-m\delta r,t\right)
F\left(  f\left(  r-m\delta r,t\right)  ,f\left(  r,t\right)  \right)
-f\left(  r,t\right)  F\left(  f\left(  r,t\right)  ,f\left(  r+m\delta
r,t\right)  \right)  \right] \nonumber\\
&  +\sum_{m=-\infty}^{\infty}p_{m} [ f\left(  r-m\delta r,t\right)  F\left(
f\left(  r-m\delta r,t\right)  ,f\left(  r,t\right)  \right)  H_{r-m,r}%
\nonumber\\
&  \;\;\;\;\;\;\;\;\;\;\;\;\;\;\;\;\;\;\;\;\;\;\;\;\; - f\left(  r,t\right)
F\left(  f\left(  r,t\right)  ,f\left(  r+m\delta r,t\right)  \right)
H_{r,r+m} ] \,.
\end{align}
The last term on the r.h.s. is
\begin{align}
&  f\left(  r,t\right)  F\left(  f\left(  r,t\right)  ,f\left(  r,t\right)
\right)  \sum_{m=-\infty}^{\infty}p_{m}\left[  H_{r-m,r}-H_{r,r+m}\right]
\nonumber\\
&  +\delta r\left(  \partial_{r}f\right)  \sum_{m=-\infty}^{\infty}%
mp_{m}\left[  -\frac{dxF}{dx}H_{r-m,r}-\frac{dxF}{dy}H_{r,r+m}\right]
+...\nonumber\\
&  =f\left(  r,t\right)  F\left(  f\left(  r,t\right)  ,f\left(  r,t\right)
\right)  \times\nonumber\\
&  \;\;\;\;\;\;\;\;\; \times\sum_{m=-\infty}^{\infty}\frac{1}{2}\left(
m\delta r\right)  ^{2}p_{m}\left[
\begin{array}
[c]{c}%
\left(  \frac{h_{yy}}{h}-2\frac{h_{x}h_{y}}{h^{2}}+2\frac{h_{x}^{2}}{h^{2}%
}-\frac{h_{xx}}{h}\right) \\
-\left(  \frac{h_{xx}}{h}-\frac{2h_{x}h_{y}}{h^{2}}+\frac{2h_{y}^{2}}{h^{2}%
}-\frac{h_{yy}}{h}\right)
\end{array}
\right]  \Theta\left(  m\left(  h_{y}-h_{x}\right)  \right) \nonumber\\
&  +\left(  \delta r\right)  ^{2}\left(  \partial_{r}f\right)  \sum
_{m=-\infty}^{\infty}m^{2}p_{m}\left[  -\frac{dxF}{dx}\left(  \frac
{h_{x}-h_{y}}{h}\right)  -\frac{dxF}{dy}\left(  \frac{h_{x}-h_{y}}{h}\right)
\right]  \Theta\left(  m\left(  f_{y}-f_{x}\right)  \right) \nonumber\\
&  =\left(  \delta r\right)  ^{2}f\left(  r,t\right)  F\left(  f\left(
r,t\right)  ,f\left(  r,t\right)  \right)  \left(  \frac{h_{yy}}{h}%
-\frac{h_{y}^{2}}{h^{2}}+\frac{h_{x}^{2}}{h^{2}}-\frac{h_{xx}}{h}\right)
\sum_{m=-\infty}^{\infty}m^{2}p_{m}\Theta\left(  m\left(  f_{y}-f_{x}\right)
\right) \nonumber\\
&  +\left(  \delta r\right)  ^{2}\left(  \partial_{r}f\right)  \left(
\frac{h_{x}-h_{y}}{h}\right)  \left[  -\frac{dxF}{dx}-\frac{dxF}{dy}\right]
\sum_{m=-\infty}^{\infty}m^{2}p_{m}\Theta\left(  m\left(  f_{y}-f_{x}\right)
\right) \nonumber\\
&  =\left(  \delta r\right)  ^{2}\left\{
\begin{array}
[c]{c}%
f\left(  r,t\right)  F\left(  f\left(  r,t\right)  ,f\left(  r,t\right)
\right)  \left(  \frac{d^{2}\ln h}{dy^{2}}-\frac{d^{2}\ln h}{dx^{2}}\right) \\
+\left(  \partial_{r}f\right)  \left[  \frac{dxF}{dx}+\frac{dxF}{dy}\right]
\left(  \frac{d\ln h}{dy}-\frac{d\ln h}{dx}\right)
\end{array}
\right\}  \sum_{m=-\infty}^{\infty}m^{2}p_{m}\Theta\left(  m\left(
f_{y}-f_{x}\right)  \right) \nonumber\\
&  =\frac{\partial}{\partial r}\left\{  f\left(  r,t\right)  F\left(  f\left(
r,t\right)  ,f\left(  r,t\right)  \right)  \left(  \frac{d\ln h}{dy}%
-\frac{d\ln h}{dx}\right)  \right\}  \left(  \delta r\right)  ^{2}%
\sum_{m=-\infty}^{\infty}m^{2}p_{m}\Theta\left(  m\left(  f_{y}-f_{x}\right)
\right) \nonumber
\end{align}
Since this term only contributes to the master equation at order $\left(
\delta r\right)  ^{2}$, it is easy to see that the complete Fokker-Planck
equation now reads
\begin{align}
&  \partial_{t}f+C\partial_{r}\left(  yF\left(  y\right)  \right)  _{f}
=D^{\prime}\left(  r\right)  \partial_{r}\left\{  f\left(  r,t\right)
F\left(  f\left(  r,t\right)  ,f\left(  r,t\right)  \right)  \left(
\frac{d\ln h}{dy}-\frac{d\ln h}{dx}\right)  _{x=y=r}\right\} \nonumber\\
&  +D\partial_{r}\left(  F\left(  y\right)  -y\frac{\partial F\left(
y\right)  }{\partial y}\right)  _{f}\partial_{r}f -\frac{1}{2}\delta
tC^{2}\partial_{r}\left(  F\left(  y\right)  -y\frac{\partial F\left(
y\right)  }{\partial y}+\left(  \frac{\partial yF\left(  y\right)  }{\partial
y}\right)  ^{2}\right)  _{f}\partial_{r}f \,,
\end{align}
where
\begin{equation}
D^{\prime}\left(  r\right)  =\frac{\left(  \delta r\right)  ^{2}}{\delta
t}\sum_{m=-\infty}^{\infty}m^{2}p_{m}\Theta\left(  m\left(  h_{y}%
-h_{x}\right)  \right)  \,.
\end{equation}
Note that in the case of symmetric elementary probabilities%
\begin{align}
\frac{\left(  \delta r\right)  ^{2}}{\delta t}  &  \sum_{m=-\infty}^{\infty
}m^{2}p_{m}\Theta\left(  m\left(  f_{y}-f_{x}\right)  \right) \nonumber\\
&  =\frac{\left(  \delta r\right)  ^{2}}{\delta t}\sum_{m>0}^{\infty}%
m^{2}\left(  p_{m}\Theta\left(  m\left(  f_{y}-f_{x}\right)  \right)
+p_{-m}\Theta\left(  -m\left(  h_{y}-h_{x}\right)  \right)  \right)
\nonumber\\
&  =\frac{\left(  \delta r\right)  ^{2}}{\delta t}\sum_{m>0}^{\infty}%
m^{2}p_{m}\left(  \Theta\left(  m\left(  f_{y}-f_{x}\right)  \right)
+\Theta\left(  -m\left(  h_{y}-h_{x}\right)  \right)  \right) \nonumber\\
&  =\frac{\left(  \delta r\right)  ^{2}}{\delta t}\sum_{m>0}^{\infty}%
m^{2}p_{m} =\frac{1}{2}\frac{\left(  \delta r\right)  ^{2}}{\delta t}%
\sum_{m=-\infty}^{\infty}m^{2}p_{m} = \overline D \,.
\end{align}

The final form of the advection-diffusion equation can be clarified. Writing
it as
\begin{align}
\partial_{t}f+C\partial_{r}\left(  yF\left(  y\right)  \right)  _{f}  &
=D^{\prime}\left(  r\right)  \frac{\partial}{\partial r}\left\{
\frac{fF\left(  f,f\right)  }{h\left(  r,r\right)  }\left(  \frac{\partial
h}{\partial y}-\frac{\partial h}{\partial x}\right)  _{x=y=r}\right\}
\nonumber\\
&  +D\partial_{r}\left(  F\left(  y\right)  -y\frac{\partial F\left(
y\right)  }{\partial y}\right)  _{f}\partial_{r}f\nonumber\\
&  -\frac{1}{2}\delta tC^{2}\partial_{r}\left(  F\left(  y\right)
-y\frac{\partial F\left(  y\right)  }{\partial y}+\left(  \frac{\partial
yF\left(  y\right)  }{\partial y}\right)  ^{2}\right)  _{f}\partial_{r}f \,,
\end{align}
and noting that%
\begin{align}
\frac{\partial}{\partial x}h\left(  x,y\right)   &  =\frac{\partial}{\partial
x}\Phi\left(  V\left(  x\right)  \right)  F\left(  \Phi\left(  V\left(
x\right)  \right)  ,\Phi\left(  V\left(  y\right)  \right)  \right)
\nonumber\\
&  =\frac{\partial\Phi}{\partial x}\left(  \frac{\partial xF\left(
x,y\right)  }{\partial x}\right)  _{\Phi} \,,
\end{align}
gives%
\begin{align}
\partial_{t}f  &  +C\, \partial_{r}\left(  yF\left(  y\right)  \right)  _{f}
=D^{\prime}\left(  r\right)  \frac{\partial}{\partial r}\left[  \frac
{fF\left(  f,f\right)  }{\Phi F\left(  \Phi,\Phi\right)  }\left(
\frac{\partial xF\left(  x,y\right)  }{\partial y}-\frac{\partial xF\left(
x,y\right)  }{\partial x}\right)  _{\Phi}\partial_{r}\Phi\right] \nonumber\\
&  + \partial_{r} \left[  D \left(  F\left(  y\right)  -y\frac{\partial
F\left(  y\right)  }{\partial y}\right)  _{f} - \frac{\delta t}{2}\,
C^{2}\partial_{r}\left(  F\left(  y\right)  -y\frac{\partial F\left(
y\right)  }{\partial y}+\left(  \frac{\partial yF\left(  y\right)  }{\partial
y}\right)  ^{2}\right)  _{f} \right]  \partial_{r}f \,. \label{A-D_eq}%
\end{align}

The local equilibrium result can also be easily deduced. It corresponds to
taking%
\begin{equation}
h\left(  x,y\right)  =\exp\left(  -\beta V\left(  x\right)  \right)
\,,\;\;\;{\mbox{i.e.}}\;\;\;\; \frac{\partial}{\partial x}h\left(  x,y\right)
=-\beta\, \frac{\partial V\left(  x\right)  }{\partial x}\exp\left(  -\beta
V\left(  x\right)  \right)  \,.
\end{equation}
Noting that the derivative of $D^{\prime}(r)$ produces a term of the form
$x\,\delta\left(  x\right)  $ which vanishes, $D^{\prime}(r)$ can be taken
under the derivative $\partial_{r}$ in (\ref{A-D_eq}), and in the local
equilibrium case
\begin{equation}
D^{\prime}\left(  r\right)  \frac{\partial}{\partial r}\left(  \beta
\frac{\partial V\left(  r\right)  }{\partial r}fF\left(  f,f\right)  \right)
=\frac{\partial}{\partial r}\left(  D^{\prime}\left(  r\right)  \beta
\frac{\partial V\left(  r\right)  }{\partial r}fF\left(  f,f\right)  \right)
\,.
\end{equation}
The resulting generalized Fokker-Planck equation reads
\begin{align}
\partial_{t}f  &  + \partial_{r} \left[  C \left(  yF\left(  y\right)
\right)  _{f} - D^{\prime} \left(  r\right)  \left(  \beta\frac{\partial V
\left(  r\right)  }{\partial r} f F\left(  f,f\right)  \right)  \right]
=\nonumber\\
&  + \partial_{r} \left[  D \left(  F\left(  y\right)  -y\frac{\partial
F\left(  y\right)  }{\partial y}\right)  _{f} -\frac{\delta t}{2}C^{2} \left(
F\left(  y\right)  -y\frac{\partial F\left(  y\right)  }{\partial y}+\left(
\frac{\partial yF\left(  y\right)  }{\partial y}\right)  ^{2}\right)  _{f}
\right]  \partial_{r}f \,.
\end{align}

\bibliographystyle{apsrev}
\bibliography{./generalized_diffusion}

\bigskip
\end{document}